\documentclass[preprint,12pt,3p]{elsarticle}

\usepackage{amssymb}
\usepackage{macros}

\usepackage{booktabs}
\usepackage{threeparttable}
\usepackage{amsmath}
\usepackage{amsthm}

\usepackage{lineno}
\usepackage{amsfonts}

\usepackage{bussproofs}
\usepackage{amsmath}
\usepackage{xcolor}
\usepackage{multirow}
\usepackage{makecell}
\usepackage{lscape}
\usepackage{stmaryrd}
\usepackage{graphicx}
\usepackage{comment}
\usepackage{mdframed}
\usepackage{enumerate}
\usepackage{float}

\usepackage{thmtools}
\usepackage{thm-restate}

\usepackage[bibliography=common, appendix=inline]{apxproof}

\usepackage{tikz}

\newtheoremrep{defn}[thm]{Definition}
\newtheoremrep{theoremapx}[thm]{Theorem}
\newtheoremrep{propositionapx}[thm]{Proposition}
\newtheoremrep{lemmaapx}[thm]{Lemma}
\newtheoremrep{corollaryapx}[thm]{Corollary}
\newtheoremrep{remarkapx}[thm]{Remark}

\usepackage{refcount}

\usepackage{hyperref}

\journal{arXiv}

\begin{document}

\begin{frontmatter}



\title{
Deducibility in the
full Lambek calculus with weakening
is 
$\UniHAck$-complete}


\author[inst1]{Vitor Greati}

\affiliation[inst1]{organization={Bernoulli Institute, University of Groningen},
            country={Nijenborgh 4, NL-9747 AG Groningen, Netherlands}}

\affiliation[inst2]{organization={CogniGron, University of Groningen},
            country={Nijenborgh 4, NL-9747 AG Groningen, Netherlands}}

\author[inst1,inst2]{Revantha Ramanayake\footnote{The financial support of the CogniGron research center, the Ubbo Emmius Funds (University of Groningen), and FWF project P33548 is acknowledged.}}


\begin{abstract}
\sloppy We show that 
the problem of deciding the consequence relation (deducibility) for
the full Lambek calculus with weakening ($\UniFLExtSCalc{\UniWProp}{}$)
is complete
for the class
$\UniHAck$ 
of hyper-Ackermannian problems (i.e., $\UniFGHProbOneAppLevel{\omega^\omega}$
in the ordinal-indexed hierarchy of fast-growing complexity classes).
 Provability was already known to be \UniCompClass{PSpace}-complete.
We prove that deducibility is $\UniHAck$-complete even for the multiplicative fragment.
Lower bounds are obtained via a novel reduction from reachability
in lossy channel systems. Upper bounds
are obtained via structural proof theory 
(forward proof search over sequent calculi)
and 
well-quasi-order theory (length theorems for Higman's ordering on words over a finite alphabet).

\end{abstract}



\begin{keyword}
substructural logics \sep%
hyper-Ackermannian problems \sep%
Lambek calculus \sep%
Higman's Lemma \sep%
well-quasi-order theory \sep%
sequent calculi
\end{keyword}

\end{frontmatter}


\section{Introduction}
\label{sec:introduction}
Substructural logics are obtained from 
intuitionistic and classical logic by
removing familiar structural properties like
exchange (e), contraction (c)
and weakening (w), and extending with further axioms and language connectives. A prominent example is the family of \emph{linear logics} where a propositional substructural base is extended with modalities (``exponentials") for weakening and contraction.
The absence of structural properties leads to highly expressive and complex logics, and this is evident even in the most fundamental of substructural logics (see Table~\ref{tab:comp-status}), namely the extensions of the full Lambek calculus $\UniFLExtSCalc{}{}$ by subsets of $\{\mathrm{(e),(c),(w)}\}$. A significant gap in this table persisted at $\UniFLExtSCalc{\UniWProp}{}$ since the deducibility problem (provability of a formula from a finite set of formulas) was, thus far, only known to be decidable, a result obtained by Blok and van Alten~\cite{blokalten2005} via the finite embeddability property for $\UniAlgebraClass{FL_{\mathsf w}}$-algebras, the algebraic semantics of $\UniFLExtLogic{\UniWProp}{}$. We fill this gap by establishing completeness with respect to the fast-growing hyper-Ackermannian complexity class $\UniHAck$, placing $\UniFLExtSCalc{\UniWProp}{}$ as the hardest for deducibility among the decidable basic substructural logics (taking the crown from $\UniFLExtLogic{\UniEProp\UniCProp}{}$, where deducibility has Ackermannian complexity).
Let $\UniFLExtSCalc{\UniWProp}{\Sigma}$ denote the restriction of $\UniFLExtSCalc{\UniWProp}{}$ to connectives in $\Sigma$.



\begin{restatable}[Main theorem]{theoremapx}{maintheorem}
\label{the:main}
        For $\{ \fus,\ld,0,1 \} \subseteq \Sigma \subseteq \UniSigFL{}$,
            deducibility in 
        $\UniFLExtSCalc{\UniWProp}{\Sigma}$
        is $\UniHAck$-complete.
            In particular, 
            $\UniFLExtLogic{\UniWProp}{}$
            and its
            multiplicative fragment
            $\UniLExtLogic{\UniWProp}{}$
            are $\UniHAck$-complete.
\end{restatable}

\begin{table}[tbh]
    \footnotesize
    \centering
    \begin{tabular}{l|l|l}
        \toprule
         Logic & Deducibility & Provability \\
         \midrule
         $\UniFLExtLogic{}{}$& Undecidable~\cite{Jipsen2002} & \UniCompClass{PSpace}-complete~\cite{horcik2011} \\
         $\UniFLExtLogic{\UniEProp}{}$& Undecidable~\cite{lincoln1992} & \UniCompClass{PSpace}-complete~\cite{horcik2011} \\
         $\UniFLExtLogic{\UniEProp\UniCProp}{}$& \UniCompClass{Ack}-complete~\cite{urquhart1999} & \UniCompClass{Ack}-complete~\cite{urquhart1999} \\
         $\UniLExtSCalc{\UniEProp\UniCProp}{}$& \UniCompClass{2-ExpTime}-complete~\cite{schmitz2011} & \UniCompClass{2-ExpTime}-complete~\cite{schmitz2011}\\
         $\UniFLExtLogic{\UniEProp\UniWProp}{}$& \UniCompClass{Tower}-complete~\cite{tanaka2022} & \UniCompClass{PSpace}-complete~\cite{horcik2011}\\
         $\UniFLExtLogic{\UniCProp}{}$& Undecidable~\cite{horcik2016} & Undecidable~\cite{horcik2016}\\
         $\UniFLExtLogic{\UniWProp}{}$& Decidable~\cite{blokalten2005}. Complexity: \textbf{this paper} & \UniCompClass{PSpace}-complete~\cite{horcik2011}\\
         $\UniLExtLogic{\UniWProp}{}$& Decidable~\cite{galatosono2010}. Complexity: \textbf{this paper} & \UniCompClass{PSpace}~\cite{horcik2011}\\
         \bottomrule
    \end{tabular}
    \caption{The computational status of the basic substructural
    logics.}
    \label{tab:comp-status}
\end{table}

The above statement references
Schmitz's ordinal-indexed \emph{fast-growing complexity classes}
$\{ \UniFGHProbOneAppLevel{\alpha} \}_{\alpha}$~\cite{schmitz2016hierar}. 
Important members of this family are
$\UniFGHProbOneAppLevel{2}$, problems solvable in
time expressed by an elementary function
(such as polynomial and exponential);
$\UniFGHProbOneAppLevel{3}$, the first class
of non-elementary problems, known as $\UniCompClass{Tower}$;
$\UniFGHProbOneAppLevel{\omega}$, the first class of
non-primitive recursive problems, known as $\UniCompClass{Ack}$
(short for `Ackermann'); 
and $\UniFGHProbOneAppLevel{\omega^\omega}$,
the first class of non-multiply recursive problems.
The complexity of the latter cannot be expressed by
a function defined by multiple nested recursion~\cite[\S 10]{roszapeter1967}, and it is known as $\UniHAck{}$ (`hyper-Ackermann').

\paragraph{An overview of the lower bound argument}
Inspired by Urquhart's~\cite{urquhart1999} lower bound for $\UniFLExtLogic{\UniEProp\UniCProp}{}$ where \emph{acceptance} for expansive additive counter machines is reduced to deducibility, we reduce \emph{reachability} 
for \emph{lossy channel systems} (LCS) to deducibility in $\UniFLExtLogic{\UniWProp}{}$.

An LCS is a computational model whose configurations are tuples $(q, u_1,\ldots,u_n)$ where $q$ is the state and $u_k$ is the $k^{\text{th}}$ channel's contents i.e., some word over a finite alphabet $\UniMessAlphabet$. Its instructions take the following form: \emph{read} letter $\UniLetter{}$ from the front of channel~$k$ in state $q_i$, remove it and transit to state $q_j$; \emph{write} letter $\UniLetter{}$ to the back of channel $k$ in state $q_i$ and transit to state $q_j$; perform a \emph{lossy} step by deleting some letters from any channel (the state remains unchanged). Chambart and Schnoebelen~\cite{chambart2008} show that deciding reachability of a configuration~$v$ from a configuration $u$ is $\UniHAck$-complete.

We code the set $\UniInstructionSet$ of instructions of an LCS $\UniLcsA$ as a finite set $\UniTheoryLCS{\UniLcsA}$ of sequents, and simulate read and write instructions by cuts with these sequents, and lossy steps by left weakening. For instance, reachability of $(q_2,\mathtt a, \mathtt b)$ from $(q_1,\mathtt a \mathtt a, \mathtt b)$ in $\UniLcsA$ is coded as deducibility of the following sequent from $\UniTheoryLCS{\UniLcsA}$:
\[
\UniSequent{
Q_1,\UniSepEnc{s}{1},A,A,\UniSepEnc{e}{1},\UniSepEnc{s}{2},B,\UniSepEnc{e}{2}
}
{
Q_2\fus(\UniSepEnc{s}{1}\fus (A\fus(\UniSepEnc{e}{1}\fus(\UniSepEnc{s}{2}\fus (B\fus\UniSepEnc{e}{2})))))
}
\]
Reading upwards from the endsequent, the following deduction
in $\UniFLExtLogic{\UniWProp}{}$ 
simulates an instruction
$(q_i,c_1,\mathtt a,?,q_j)\in\UniInstructionSet$; i.e., read $\mathtt a$ from the front of channel $c_1$ in state $q_i$, remove it, then transit to $q_j$.
\begin{center}
\AxiomC{sequent in $\UniTheoryLCS{\UniLcsA}$ coding instruction}
\noLine
\UnaryInfC{$\UniSequent{Q_i,\UniSepEnc{s}{1}, A}{Q_j\fus \UniSepEnc{s}{1}}$}
\AxiomC{$\cdots$}
\noLine
\UnaryInfC{$\UniSequent{Q_j, \UniSepEnc{s}{1},B,C,\UniSepEnc{e}{1}}{v}$}
\RightLabel{$\UniRuleSide{L}{\fus}$}
\UnaryInfC{$\UniSequent{Q_j \fus \UniSepEnc{s}{1},B,C,\UniSepEnc{e}{1}}{v}$}
\RightLabel{(cut)}
\BinaryInfC{$\UniSequent{Q_i,\UniSepEnc{s}{1},A,B,C,\UniSepEnc{e}{1}}{v}$}
\DisplayProof
\end{center}
Every computation in the LCS is shown to correspond to a deduction of its reachability encoding. The converse direction is far more challenging, requiring a careful analysis of deductions in $\UniFLExtLogic{\UniWProp}{}$. The starting point is a cut-elimination style argument that establishes a normal form (of ``standard deductions") where the left premise of every cut is an element of the theory. The computation is read off the standard deduction, starting at the endsequent.

\paragraph{The upper bound argument} We define a forward proof procedure extending the construction in Balasubramanian et al.~\cite{BalLanRam21LICS}. The idea is to start from the set $\UniDerivSet_{0}$ of initial sequent instances and form an increasing chain $\UniDerivSet_{0}\subset \UniDerivSet_{1}\subset\ldots$ of finite sets of deducible sequents (each step corresponds to limited weakening and then a single rule application), taking care that each $\UniDerivSet_{i}$ contains only minimal elements under the Higman ordering on words (omit words that can be obtained from another by inserting letters); the ordering is a well-quasi-order (wqo)~\cite{higman1952} so the chain must be finite. By construction, the bad sequence underlying the chain is controlled (no element in $\UniDerivSet_{i+1}$ is much larger than some element in $\UniDerivSet_{i}$), and the complexity is dominated by the maximum length of a controlled bad sequence. This is a function in $\UniHAck{}$ as shown by Schmitz and Schnoebelen~\cite{schmitz2011}. The argument applies to infinitely-many structural rule extensions of $\UniFLExtLogic{\UniWProp}{}$.

\paragraph{Contributions}
We show that
(1)
        sequent-deducibility
        in fragments of
        $\UniFLExtSCalc{\UniWProp}{}$
        containing $\fus$
        is $\UniHAck$-hard
        and that
        (2) deducibility for any $\mathcal{N}_2$-analytic structural rule extension~\cite{CiaGalTer17} of any fragment of
        $\UniFLExtSCalc{\UniWProp}{}$ is in $\UniHAck$.
Our main theorem follows from
(1) and~(2). Algebraizability of $\UniFLExtSCalc{\UniWProp}{}$ implies that (3)~the word problem 
and the quasiequational theory
of the variety of
integral zero-bounded full Lambek algebras are
$\UniHAck$-complete.

\section{Noncommutative substructural logics with weakening}
\label{sec:substructural-logics}

Let $\Sigma$ be a propositional signature
and fix a countable set $\UniPropSet$ of 
\emph{propositional variables}.
Let $\UniLangSet{\Sigma}{\UniPropSet}$ denote
the set of \emph{formulas} over $\Sigma$
generated by $\UniPropSet$.
We consider subsignatures of
$\UniSigFL{} \UniSymbDef 
\{ 0, 1, \top, \bot, \fus, \land, \lor, \rd, \ld\}$,
where $0, 1, \top$ and $\bot$ are constants
and the other connectives are binary.
A subsignature is specified as
$\UniSigFL{\UniConA_1\cdots\UniConA_k} \UniSymbDef \{ \UniConA_1,\ldots,\UniConA_k \} \subseteq \UniSigFL{}$. For example, $\UniSigFL{\fus\ld} = \{ \fus, \ld \}$.
Sequences (or lists) of formulas
will be written without
delimiters, with commas separating the formulas. 
We write
$\bigotimes \UniList{\UniFmA_1,\ldots,\UniFmA_m}$ as abbreviation for $\UniFmA_1 \fus (\UniFmA_2 \fus ( \ldots ( \UniFmA_{m-1} \fus \UniFmA_m )\ldots))$, where $m \geq 1$
and $\UniFmA_1,\ldots,\UniFmA_m$
is a list of formulas.
If the constant $1$ is present in the language,
we allow $m=0$ (empty fusion), the result being $1$.

A \emph{sequent over $\UniLangSet{\Sigma}{\UniPropSet}$} is an expression
$\UniSequent{\UniMSetFmA}{\UniMSetSucA}$,
where $\UniMSetFmA$ is a finite sequence of
formulas and $\UniMSetSucA$
is an empty sequence or a sequence with a single
formula.
Given a set $\mathcal{T}$ of sequents,
let $\UniSubf{\mathcal{T}}$ denote the set of subformulas of the formulas appearing in it.
We assume familiarity
with the sequent calculus
(see e.g., \cite[Sec. 2.1.3]{GalJipKowOno07}): 
a \emph{sequent calculus $\UniSeqCalcA$ over $\UniLangSet{\Sigma}{\UniPropSet}$} is a collection of
schematic sequent rules. \emph{Instantiations} of
their schematic variables
lead to \emph{rule instances}.
\emph{Derivations} in $\UniSeqCalcA$ are finite rooted trees labelled with sequents over $\UniLangSet{\Sigma}{\UniPropSet}$
such that a node and its children are the conclusion and premises of a rule instance.
For a finite set $\mathcal{T}$ of sequents, a \emph{$\mathcal{T}$-deduction}
in $\UniSeqCalcA$
is a derivation in which every leaf is a rule
with no premises (\emph{axiom} or \emph{initial sequent})
or a sequent in $\mathcal{T}$.
A \emph{proof} in $\UniSeqCalcA$ is a
$\varnothing$-deduction in $\UniSeqCalcA$.
Any $\UniSeqCalcA' \subseteq \UniSeqCalcA$
is called a \emph{subcalculus} of $\UniSeqCalcA$,
and any $\UniSeqCalcA'' \supseteq \UniSeqCalcA$
is a \emph{rule extension} of $\UniSeqCalcA$.
\emph{Structural rule extensions} are extensions
by \emph{structural rules} (rules without logical connectives).

Subcalculi of the sequent calculus $\UniFLExtSCalc{\mathbf{w}}{}$
displayed in Fig.~\ref{figure-HFLec} (\emph{full Lambek calculus with weakening}) are induced by a subsignature of $\UniSigFL{}$
as follows.
Let $\UniFLExtSCalc{\mathbf{w}}{\UniConA_1\cdots\UniConA_k}$
be the sequent calculus consisting of
rules that mention no connective
(i.e., rule $\UniRule{(id)}$ and structural rules),
or
that mentions a connective in $\{\UniConA_1,\ldots,\UniConA_k\}$.
For instance, 
$\UniFLExtSCalc{\mathbf{w}}{\fus\ld}$
is the calculus 
over $\UniLangSet{\UniSigFL{\fus\ld}}{\UniPropSet}$
consisting of the axioms, all structural rules
and the rules mentioning $\fus$ and $\ld$.


A sequent calculus $\UniSeqCalcA$ over $\UniLangSet{\Sigma}{\UniPropSet}$
canonically
determines
two \emph{consequence relations}, one
over sequents and another over formulas.
The \emph{sequent-deduction relation} $\UniHyperDerivRel{\UniSeqCalcA}$ 
is
such that, for all finite sets $\mathcal{T}\cup\{\UniSequentA\}$ of sequents over $\UniLangSet{\Sigma}{\UniPropSet}$,
$\mathcal{T} \UniHyperDerivRel{\UniSeqCalcA} \UniSequentA$
iff there is a $\mathcal{T}$-deduction
of $\UniSequentA$ in $\UniSeqCalcA$.
The \emph{formula-deduction relation}
$\vdash_{\UniSeqCalcA}$ 
is such that, for all finite sets $\UniSetFmA\cup\{\UniFmA\}$ of
formulas over $\UniLangSet{\Sigma}{\UniPropSet}$,
$\UniSetFmA \vdash_{\UniSeqCalcA} \UniFmA$
iff
there is a $\{ \UniSequent{}{\UniFmB} \mid \UniFmB \in \UniSetFmA \}$-deduction of $\UniSequent{}{\UniFmA}$
in $\UniSeqCalcA$.
Given a sequent calculus $\UniSeqCalcA$,
the problem of deciding $\UniHyperDerivRel{\UniSeqCalcA}$
is called \emph{sequent-deducibility}
and the problem of deciding 
$\vdash_{\UniSeqCalcA}$ is called
\emph{formula-deducibility}
or simply \emph{deducibility}.
We know that
$\vdash_{\UniFLExtSCalc{\mathbf{w}}{}}$ is the consequence relation
of the full Lambek logic with weakening
(see~\cite[2.1.4]{GalJipKowOno07} for the Hilbert-calculus presentation and residuated lattices semantics of these logics).
The reader is also referred to~\cite[2.1.4]{GalJipKowOno07} for the notion of
\emph{axiomatic extensions} of $\vdash_{\UniFLExtSCalc{\mathbf{w}}{}}$.

%

\sloppy
The following lemma says that
$\UniHyperDerivRel{\UniSeqCalcA}$
reduces to
$\vdash_{\UniSeqCalcA}$ when
$\UniSeqCalcA$ is a subcalculus of 
$\UniFLExtSCalc{\mathbf{w}}{}$
containing the rules for $\fus,\ld,0$ and $1$.
For that,
consider the translation $\UniSeqToFormEnc{(\cdot)}$
of sequents over $\UniLangSet{\Sigma}{\UniPropSet}$ 
into formulas over $\UniLangSet{\Sigma \cup \{ 0,1,\fus,\ld \}}{\UniPropSet}$ defined
$\UniSeqToFormEnc{(\UniSequent{\UniMSetFmA}{\UniMSetSucA})} \UniSymbDef \bigotimes\UniMSetFmA \ld \UniMSetSucA^\star$,
where $\UniMSetSucA^\star$ is $\UniFmA$
if $\UniMSetSucA = \UniFmA$ and $0$ if $\UniMSetSucA$
is empty.

\begin{lemmaapxrep}\label{lem:der-seq-deducibility}
    Let $\Sigma \supseteq \{ \fus,\ld,0,1 \}$
    and
    $\UniSeqCalcA \UniSymbDef
\UniFLExtSCalc{\mathbf{w}}{\Sigma}$.
For all finite sets of sequents $\mathcal{T}$
    and sequents $s$
    over the language of $\UniSeqCalcA$,
    $
    \mathcal{T} \UniHyperDerivRel{\UniSeqCalcA} \UniSequentA
    \text{ iff }
    \{ \UniSeqToFormEnc{\UniSequentB} \mid \UniSequentB \in \mathcal{T} \}
    \vdash_{\UniSeqCalcA} \UniSeqToFormEnc{s}.
    $
\end{lemmaapxrep}
\begin{proof}
    From left to right,
    let $\delta$ be a deduction
    witnessing 
    $\mathcal{T} \UniHyperDerivRel{\UniSeqCalcA} \UniSequentA$.
    We show how to convert $\delta$ into a witness of
    $\{ \UniSeqToFormEnc{\UniSequentB} \mid \UniSequentB \in \mathcal{T} \}
    \vdash_{\UniSeqCalcA} \UniSeqToFormEnc{s}$
    by structural induction.
    In the base case,
    $\UniSequentA$
    is either
    the result of
    an axiomatic rule
    or an element of $\mathcal{T}$.
    To translate $\UniSequentB \in \mathcal{T}$ into $\UniSeqToFormEnc{\UniSequentB}$,
    repeatedly use the rules
    $\UniRuleSide{L}{\fus}$ and $\UniRuleSide{R}{\ld}$, and
    $\UniRuleSide{L}{1}$ and $\UniRuleSide{R}{0}$ if required.
    In the inductive step,
    assume $\UniSequent{\UniMSetFmA}{\UniMSetSucA}$ was derived by an application of a $k$-ary
    rule $\UniRuleA$,
    whose premises are
    the sequents
    $\UniSequent{\UniMSetFmA_1}{\UniMSetSucA_1}, \ldots,
    \UniSequent{\UniMSetFmA_k}{\UniMSetSucA_k}$.
    Note first that the rule
        \AxiomC{$\UniSequent{}{\bigotimes\UniMSetFmA \ld \UniMSetSucA}$}
    \RightLabel{(FS)}
    \UnaryInfC{$\UniSequent{\UniMSetFmA}{\UniMSetSucA}$}
    \DisplayProof
    is derivable in presence of cut.
    By IH,
    the sequents
    $\UniSequent{}{\bigotimes\UniMSetFmA_1 \ld \UniMSetSucA_1^\star}, \ldots,
    \UniSequent{}{\bigotimes\UniMSetFmA_k \ld \UniMSetSucA_k^\star}$ are
    provable from $\mathcal{T}^\to$.
    Apply to these premises
    the derived rule $\UniRule{(FS)}$
    thus obtaining the premises 
    $\UniSequent{\UniMSetFmA_1}{\UniMSetSucA_1^\star}, \ldots,
    \UniSequent{\UniMSetFmA_k}{\UniMSetSucA_k^\star}$
    (maybe cuts with $\UniSequent{}{1}$
    will be necessary to recover
    empty antecedents).
    Use cuts with
    $\UniSequent{0}{}$
    whenever necessary to obtain from those the premises $\UniSequent{\UniMSetFmA_1}{\UniMSetSucA_1}, \ldots,
    \UniSequent{\UniMSetFmA_k}{\UniMSetSucA_k}$.
    Apply the same rule 
    $\UniRuleA$
    to the latter,
    obtaining $\UniSequent{\UniMSetFmA}{\UniMSetSucA}$, from
    which $\UniSequent{}{\bigotimes \UniMSetFmA \ld \UniMSetSucA^\star}$
    is easily derivable (if $\UniMSetFmA$ is empty, use $\UniRuleSide{L}{1}$), and we are done.

    From right to left,
    first transform
    all the leaves 
    with translated sequents from $\mathcal{T}$
    into trees with leaves being
    either instances of axioms or
    the sequents themselves.
    This is easy to do using
    rules $\UniRuleSide{R}{\ld}$, $\UniRuleSide{L}{\fus}$,
    $\UniRuleSide{L}{1}$ and $\UniRuleSide{R}{0}$.
    Then it is enough to
    transform the root 
    $\UniSeqToFormEnc{\UniSequentA}$
    into
    $\UniSequentA$
    using $\UniRule{(FS)}$ as we did in the converse direction.
    \qedhere
\end{proof}

\begin{figure}[tbh!]
    \begin{center}
    \begin{footnotesize}
    \begin{tabular}{c@{\hspace{1em}}c@{\hspace{1em}}c@{\hspace{1em}}c@{\hspace{1em}}c}
        \AxiomC{}
        \RightLabel{(id)}
        \UnaryInfC{
        $\UniSequent{\UniSchPropA}{\UniSchPropA}$
        }
        \DisplayProof
        &
        \AxiomC{}
        \RightLabel{$\UniRuleSide{L}{\bot}$}
        \UnaryInfC{
        $ 
        \UniSequent{\UniMSetFmA, \bot, \UniMSetFmB}{\UniMSetSucA}$
        }
        \DisplayProof
        &
        \AxiomC{}
        \RightLabel{$\UniRuleSide{R}{\top}$}
        \UnaryInfC{
        $   
        \UniSequent{\UniMSetFmA}{\top}$
        }
        \DisplayProof
        &
        \AxiomC{}
        \RightLabel{$\UniRuleSide{L}{0}$}
        \UnaryInfC{
        $ 
        \UniSequent{0}{}$
        }
        \DisplayProof
        &
        \AxiomC{}
        \RightLabel{$\UniRuleSide{R}{1}$}
        \UnaryInfC{
        $   
        \UniSequent{}{1}$
        }
        \DisplayProof
        \end{tabular}
\\[1em]                
        

            \begin{tabular}{r@{\hspace{2em}}r@{\hspace{2em}}r}
                \AxiomC{$\UniSequent{\UniMSetFmA,\UniMSetFmB}{\UniMSetSucA}$}
                \RightLabel{$\UniIRule$}
                \UnaryInfC{$\UniSequent{\UniMSetFmA,\UniFmA,\UniMSetFmB}{\UniMSetSucA}$}
                \DisplayProof
                &
                \AxiomC{$\UniSequent{\UniMSetFmA}{}$}
                \RightLabel{$\UniORule$}
                \UnaryInfC{$\UniSequent{\UniMSetFmA}{\UniFmA}$}
                \DisplayProof
                &
                \AxiomC{$\UniSequent{\UniMSetFmC}{\UniFmA}$}
                \AxiomC{$\UniSequent{\UniMSetFmA,\UniFmA,\UniMSetFmB}{\UniMSetSucA}$}
                \RightLabel{(cut)}
                \BinaryInfC{$\UniSequent{\UniMSetFmA,\UniMSetFmC,\UniMSetFmB}{\UniMSetSucA}$}
                \DisplayProof
            \end{tabular}
\\[1em]        


            \begin{tabular}{c@{\hspace{1.5em}}c@{\hspace{1.5em}}c}
            \AxiomC{$
            \UniSequent{\UniMSetFmA}{}
            $}
                \RightLabel{$\UniRuleSide{R}{0}$}
            \UnaryInfC{$
            \UniSequent{\UniMSetFmA}{0}$}
            \DisplayProof
&
             \AxiomC{$
            \UniSequent{\UniMSetFmA,\UniMSetFmB}{\UniMSetSucA}$}
            \RightLabel{$\UniRuleSide{L}{1}$}
            \UnaryInfC{$
            \UniSequent{\UniMSetFmA,1,\UniMSetFmB}{\UniMSetSucA}$}
            \DisplayProof 
&
            \AxiomC{$
            \UniSequent{\UniMSetFmA,\UniFmA,\UniFmB,\UniMSetFmB}{\UniMSetSucA}$}
            \RightLabel{$\UniRuleSide{L}{\fus}$}
            \UnaryInfC{$
            \UniSequent{\UniMSetFmA,\UniFmA\fus \UniFmB,
            \UniMSetFmB}{ \UniMSetSucA}$}
            \DisplayProof
            \end{tabular}
            \\[1em] 
            
            \begin{tabular}{c@{\hspace{1em}}c@{\hspace{1em}}c}
            \AxiomC{$
            \UniSequent{\UniMSetFmA}{\UniFmA}$}
            \AxiomC{$
            \UniSequent{\UniMSetFmB}{\UniFmB}$}
            \RightLabel{$\UniRuleSide{R}{\fus}$}
            \BinaryInfC{$  
            \UniSequent{\UniMSetFmA,\UniMSetFmB}{\UniFmA\fus \UniFmB}$}
            \DisplayProof
&
            \AxiomC{$ 
            \UniSequent{\UniMSetFmA,\UniFmA,\UniMSetFmB}{\UniMSetSucA}$
            }
            \AxiomC{$  
            \UniSequent{\UniMSetFmA,\UniFmB,\UniMSetFmB}{\UniMSetSucA}
            $
            }
            \RightLabel{$\UniRuleSide{L}{\lor}$}
            \BinaryInfC{$ 
            \UniSequent{\UniMSetFmA,\UniFmA\lor\UniFmB,\UniMSetFmB}{\UniMSetSucA}$}
            \DisplayProof
&
            \AxiomC{$  
            \UniSequent{\UniMSetFmA}{\UniFmA_i}
            $}
            \RightLabel{$\UniRuleSide{R}{\lor}$}
            \UnaryInfC{$
            \UniSequent{\UniMSetFmA}{\UniFmA_1\lor\UniFmA_2}$}
            \DisplayProof
            \end{tabular}
            \\[1em] 
            
\begin{tabular}{c@{\hspace{1em}}c@{\hspace{1em}}c}
            \AxiomC{$
            \UniSequent{\UniMSetFmA,\UniFmA_i,\UniMSetFmB}{\UniMSetSucA}$}
            \RightLabel{$\UniRuleSide{L}{\land}$}
            \UnaryInfC{$  
            \UniSequent{\UniMSetFmA,\UniFmA_1\land \UniFmA_2,
            \UniMSetFmB}{\UniMSetSucA}$}
            \DisplayProof
&            
            \AxiomC{$  
            \UniSequent{\UniMSetFmA}{\UniFmA}$}
            \AxiomC{$  
            \UniSequent{\UniMSetFmA}{\UniFmB}$}
            \RightLabel{$\UniRuleSide{R}{\land}$}
            \BinaryInfC{$  
            \UniSequent{\UniMSetFmA}{\UniFmA\land \UniFmB}$}
            \DisplayProof
&
            \AxiomC{$  
            \UniSequent{\UniMSetFmA}{\UniFmA}$}
            \AxiomC{$  
            \UniSequent{\UniMSetFmB,\UniFmB,\UniMSetFmC}{\UniMSetSucA}$}
            \RightLabel{$\UniRuleSide{L}{\rd}$}
            \BinaryInfC{$  
            \UniSequent{\UniMSetFmB,\UniFmB\rd\UniFmA,\UniMSetFmA,\UniMSetFmC}{\UniMSetSucA}$}
            \DisplayProof
\end{tabular}
\\[1em] 

\begin{tabular}{c@{\hspace{1em}}c@{\hspace{1em}}c}
            \AxiomC{$  
            \UniSequent{\UniMSetFmA,\UniFmA}{\UniFmB}$}
            \RightLabel{$\UniRuleSide{R}{\rd}$}
            \UnaryInfC{$  
            \UniSequent{\UniMSetFmA}{\UniFmB\rd \UniFmA}$}
            \DisplayProof
&
            \AxiomC{$  
            \UniSequent{\UniMSetFmA}{\UniFmA}$}
            \AxiomC{$  
            \UniSequent{\UniMSetFmB,\UniFmB,\UniMSetFmC}{\UniMSetSucA}$}
            \RightLabel{$\UniRuleSide{L}{\ld}$}
            \BinaryInfC{$  
            \UniSequent{\UniMSetFmB,\UniMSetFmA,\UniFmA\ld\UniFmB,\UniMSetFmC}{\UniMSetSucA}$}
            \DisplayProof
&
            \AxiomC{$  
            \UniSequent{\UniFmA,\UniMSetFmA}{\UniFmB}$}
            \RightLabel{$\UniRuleSide{R}{\ld}$}
            \UnaryInfC{$  
            \UniSequent{\UniMSetFmA}{\UniFmA \ld \UniFmB}$}
            \DisplayProof
        \end{tabular}
        \end{footnotesize}
    \end{center}
    \caption{The sequent calculus~$\UniFLExtSCalc{\mathbf{w}}{}$.}
    \label{figure-HFLec}
\end{figure}


Every $\varnothing$-deduction (i.e., every proof) in $\UniFLExtSCalc{\mathbf{w}}{\Sigma}$ can be converted to a proof without
cuts (by cut elimination~\cite[4.1.1]{GalJipKowOno07}), so these calculi satisfy the \emph{subformula property}
(every formula that occurs in the proof is a subformula of the endsequent).
However, not all cuts can be eliminated from a $\mathcal{T}$-deduction. Still, for suitable $\mathcal{T}$ identified below, the applications of cut can be restricted to obtain a generalized subformula property.

\begin{defn}\label{def:regular-theory-standard-cut}
A sequent is said to be
\emph{regular} if its antecedent
 contains only propositional variables
 {and the succedent is not empty}.
A \emph{regular theory} is a finite
collection of regular sequents.
For a regular theory $\mathcal{T}$,
a \emph{standard cut} (over $\mathcal{T}$)
has as left premise a leaf that is a sequent from
$\mathcal{T}$. A \emph{standard deduction} 
(from $\mathcal{T}$)
is
one in which all cuts are standard.
We write 
$\mathcal{T}\UniHyperStdDerivRel{\UniFLExtSCalc{\mathbf{w}}{\Sigma}} \UniSequent{\UniMSetFmA}{\UniFmA}$
to denote that $\UniSequent{\UniMSetFmA}{\UniFmA}$
has a standard deduction from $\mathcal{T}$ in 
$\UniFLExtSCalc{\mathbf{w}}{\Sigma}$.
\end{defn}


\begin{lemmaapxrep}\label{lem:cut-restriction}
    Let $\Sigma \subseteq \UniSigFL{}$.
For a regular theory $\mathcal{T}$
    and sequent $s$,
    $
    \mathcal{T} \UniHyperDerivRel{\UniFLExtSCalc{\mathbf{w}}{\Sigma}} \UniSequentA
    $
    iff
    $\mathcal{T}\UniHyperStdDerivRel{\UniFLExtSCalc{\mathbf{w}}{\Sigma}}
    \UniSequentA$.
    Moreover, every formula in a standard deduction
    in $\UniFLExtSCalc{\mathbf{w}}{\Sigma}$
    is in
    $\UniSubf{\mathcal{T} \cup \{ \UniSequentA \}}$.
\end{lemmaapxrep}
\begin{proof}
    
    We establish the claim for
    $\Sigma = \UniSigFL{}$. The reader will observe that removing rules means less cases to be checked,
    and thus the argument here applies to any $\Sigma \subseteq \UniSigFL{}$.
    Let us call the instantiation of the schematic-variable
$\UniFmA$ in an instance of the $\UniRule{(cut)}$ rule (cf. Fig.~\ref{figure-HFLec})
as the \emph{cut formula} of that instantiation.

    For first assertion, the argument 
    is based on the usual cut-elimination procedure
    for $\UniFLExtSCalc{\UniWProp}{}$~\cite[4.1.1]{GalJipKowOno07};
    the difference is that
    we now have to deal with cuts with a premise in 
    $\mathcal{T}$.
    We repeatedly eliminate topmost non-standard cuts,
    by primary induction on the grade of the cut (the length of the cut formula), and secondary induction on the cut-height (the sum of the number of sequents appearing in the proofs of the premises of the cut).

    Consider a topmost non-standard cut with cut-height $2$
    (note that $<2$ is not possible).
    The premises of the cut are
    instances of axiomatic rules
    or elements from $\mathcal{T}$.
    If no element from
    $\mathcal{T}$ 
    appears as a premise, the cut is eliminated in the usual way as for $\UniFLExtSCalc{\UniWProp}{}$. 
    If the left premise
    is from $\mathcal{T}$,
    nothing needs to be done since we admit
    standard cuts.
    If the right premise
    is from $\mathcal{T}$,
    the left premise can only
    be an instance either of
    $\UniRule{(id)}$
    or $\UniRuleSide{L}{\bot}$;
    i.e., the cut has one of
    these forms:
    
    \begin{center}
        \footnotesize
        \AxiomC{}
        \RightLabel{$\UniRule{(id)}$}
        \UnaryInfC{$\UniSequent{p}{p}$}
        \AxiomC{}
        \RightLabel{$\mathcal{T}$}
        \UnaryInfC{$\UniSequent{\UniMSetFmA_1,p,\UniMSetFmA_2}{\UniFmA}$}
        \RightLabel{$\UniRule{(cut)}$}
        \BinaryInfC{$\UniSequent{\UniMSetFmA_1,p,\UniMSetFmA_2}{\UniFmA}$}
        \DisplayProof
        \;
        \AxiomC{}
        \RightLabel{$\UniRule{(\bot L)}$}
        \UnaryInfC{$\UniSequent{\UniMSetFmB_1,\bot,\UniMSetFmB_2}{p}$}
        \AxiomC{}
        \RightLabel{$\mathcal{T}$}
        \UnaryInfC{$\UniSequent{\UniMSetFmA_1,p,\UniMSetFmA_2}{\UniFmA}$}
        \RightLabel{$\UniRule{(cut)}$}
        \BinaryInfC{$\UniSequent{\UniMSetFmA_1,\UniMSetFmB_1,\bot,\UniMSetFmB_2,\UniMSetFmA_2}{\UniFmA}$}
        \DisplayProof
    \end{center}
    
    \noindent The cut above left is clearly
    eliminable since the second premise
    matches the conclusion;
    the cut above right is replaced by an instance of 
    $\UniRuleSide{L}{\bot}$.

    Inductive step. Assume the 
    cut has cut-height $\geq 3$.
    {
    If neither of the premises of cut are
    from $\mathcal{T}$, the usual cut-reduction steps
    apply~\cite[4.1.1]{GalJipKowOno07}}.
    Else, if the topmost cut is 
    standard, nothing needs to be done
    since we admit such cuts.
The remaining case is that the right premise is from $\mathcal{T}$ and the left premise is from some other rule. It follows that the topmost non-standard cut occurs as follows.

    \begin{center}
        \AxiomC{$\delta_1$}
        \noLine
        \UnaryInfC{$\UniSequent{\UniMSetFmA}{p}$}
        \AxiomC{}
        \RightLabel{$\mathcal{T}$}
        \UnaryInfC{$\UniSequent{\UniMSetFmB_1,p,\UniMSetFmB_2}{\UniMSetSucA}$}
        \RightLabel{$\UniRule{(cut)}$}
        \BinaryInfC{$\UniSequent{\UniMSetFmB_1,\UniMSetFmA,\UniMSetFmB_2}{\UniMSetSucA}$}
        \DisplayProof
    \end{center}

    Since the cut-height 
    is $\geq 3$, we 
    have that the left
    premise is not
    an instance of an 
    axiomatic rule.
    Consider the last rule in $\delta_1$.
    We transform each case so the non-standard 
    cut is replaced with a cut that is smaller under the induction measure.
    We indicate the transformations
    with $\rightsquigarrow$.
    Since the cut formula 
    is a propositional variable (this is forced by the right premise being from the regular theory $\mathcal{T}$), the last rule in $\delta_1$ cannot be a right-introduction rule.
    We provide some representative cases.

    \begin{itemize}
        \item     
            The left premise of the cut under consideration is the conclusion of a standard cut:
        
            \begin{center}
                \AxiomC{}
                \RightLabel{$\mathcal{T}$}
                \UnaryInfC{$\UniSequent{\UniMSetFmC}{\UniFmA}$}
                \AxiomC{$\delta_1$}
                \noLine
                \UnaryInfC{$\UniSequent{\UniMSetFmA_1,\UniFmA,\UniMSetFmA_2}{p}$}
                \RightLabel{$\UniRule{(cut)}$}
                \BinaryInfC{$\UniSequent{\UniMSetFmA_1,\UniMSetFmC,\UniMSetFmA_2}{p}$}
                \AxiomC{}
                \RightLabel{$\mathcal{T}$}
                \UnaryInfC{$\UniSequent{\UniMSetFmB_1,p,\UniMSetFmB_2}{\UniMSetSucA}$}
                \RightLabel{$\UniRule{(cut)}$}
                \BinaryInfC{$\UniSequent{\UniMSetFmB_1,\UniMSetFmA_1,\UniMSetFmC,\UniMSetFmA_2,\UniMSetFmB_2}{\UniMSetSucA}$}
                \noLine
                \UnaryInfC{}
                \DisplayProof
                
                $\UniRwDown$

                \AxiomC{}
                \RightLabel{$\mathcal{T}$}
                \UnaryInfC{$\UniSequent{\UniMSetFmC}{\UniFmA}$}
                \AxiomC{$\delta_1$}
                \noLine
                \UnaryInfC{$\UniSequent{\UniMSetFmA_1,\UniFmA,\UniMSetFmA_2}{p}$}
                \AxiomC{}
                \RightLabel{$\mathcal{T}$}
                \UnaryInfC{$\UniSequent{\UniMSetFmB_1,p,\UniMSetFmB_2}{\UniMSetSucA}$}
                \RightLabel{$\UniRule{(cut)}$}
                \BinaryInfC{$\UniSequent{\UniMSetFmB_1,\UniMSetFmA_1,\UniFmA,\UniMSetFmA_2,\UniMSetFmB_2}{p}$}
                \RightLabel{$\UniRule{(cut)}$}
                \BinaryInfC{$\UniSequent{\UniMSetFmB_1,\UniMSetFmA_1,\UniMSetFmC,\UniMSetFmA_2,\UniMSetFmB_2}{\UniMSetSucA}$}
                \DisplayProof
            \end{center}

        \item $\UniRule{(L\fus)}$
            \begin{center}
                \AxiomC{$\delta_2$}
                \noLine
                \UnaryInfC{$\UniSequent{\UniMSetFmA_1,\UniFmA_1,\UniFmA_2,\UniMSetFmA_2}{p}$}
                \RightLabel{$\UniRule{(L\fus)}$}
                \UnaryInfC{$\UniSequent{\UniMSetFmA_1,\UniFmA_1\fus\UniFmA_2,\UniMSetFmA_2}{p}$}
                \AxiomC{}
                \RightLabel{$\mathcal{T}$}
                \UnaryInfC{$\UniSequent{\UniMSetFmB_1,p,\UniMSetFmB_2}{\UniMSetSucA}$}
                \RightLabel{$\UniRule{(cut)}$}
                \BinaryInfC{$\UniSequent{\UniMSetFmB_1,\UniMSetFmA_1,\UniFmA_1\fus\UniFmA_2,\UniMSetFmA_2,\UniMSetFmB_2}{\UniMSetSucA}$}
                \noLine
                \UnaryInfC{}
                \DisplayProof
                
                $\UniRwDown$
                
                \AxiomC{$\delta_2$}
                \noLine
                \UnaryInfC{$\UniSequent{\UniMSetFmA_1,\UniFmA_1,\UniFmA_2,\UniMSetFmA_2}{p}$}
                \AxiomC{}
                \RightLabel{$\mathcal{T}$}
                \UnaryInfC{$\UniSequent{\UniMSetFmB_1,p,\UniMSetFmB_2}{\UniMSetSucA}$}
                \RightLabel{$\UniRule{(cut)}$}
                \BinaryInfC{$\UniSequent{\UniMSetFmB_1,\UniMSetFmA_1,\UniFmA_1,\UniFmA_2,\UniMSetFmA_2,\UniMSetFmB_2}{\UniMSetSucA}$}
                \RightLabel{$\UniRule{(L\fus)}$}
                \UnaryInfC{$\UniSequent{\UniMSetFmB_1,\UniMSetFmA_1,\UniFmA_1\fus\UniFmA_2,\UniMSetFmA_2,\UniMSetFmB_2}{\UniMSetSucA}$}
                \DisplayProof
            \end{center}
    \end{itemize}
    
    The second assertion follows by structural induction
    on standard deductions.
    The base case is obvious.
    Inductive step. The claim is immediate for a logical rule, $\UniIRule$, and $\UniORule$, by the induction hypothesis applied to the deduction concluding the premise, and since every formula in the premise must be a subformula of a formula in the conclusion. It remains to check the case of a standard cut.
    \begin{center}
        \AxiomC{}
        \RightLabel{$\mathcal{T}$}
        \UnaryInfC{$\UniSequent{\UniMSetFmB}{\UniFmB}$}
        \AxiomC{$\delta_1$}
        \noLine
        \UnaryInfC{$\UniSequent{\UniMSetFmA_1,\UniFmB,\UniMSetFmA_2}{\UniFmA}$}
        \RightLabel{cut}
        \BinaryInfC{$\UniSequent{\UniMSetFmA_1,\UniMSetFmB,\UniMSetFmA_2}{\UniFmA}$}
        \DisplayProof
    \end{center}
    The claim follows since
    $\UniFmB$ is a formula present in one of the
    sequents in $\mathcal{T}$.
\end{proof}


The next lemma 
is a direct consequence of Lem.~\ref{lem:cut-restriction}
and
tells us
that sequent-deducibility in ${\UniFLExtSCalc{\mathbf{w}}{\Sigma'}}$
reduces to the same problem over
an expanded signature $\UniSigFL{} \supseteq \Sigma\supseteq\Sigma'$.
In other words,
$\UniHyperDerivRel{\UniFLExtSCalc{\mathbf{w}}{\Sigma}}$ is a \emph{conservative
expansion} of 
$\UniHyperDerivRel{\UniFLExtSCalc{\mathbf{w}}{\Sigma'}}$.

\begin{lemmaapxrep}\label{lem:subsig-sig-der-seq-equiv}
    Let $\Sigma' \subseteq \Sigma \subseteq \UniSigFL{}$. 
If $\mathcal{T} \cup \{\UniSequentA\}$
    are sequents over $\UniLangSet{\Sigma'}{\UniPropSet}$,
then
$
    \mathcal{T} \UniHyperDerivRel{\UniFLExtSCalc{\mathbf{w}}{\Sigma'}} \UniSequentA
    $
    iff
$    \mathcal{T}\UniHyperDerivRel{\UniFLExtSCalc{\mathbf{w}}{\Sigma}} \UniSequentA.
$
\end{lemmaapxrep}
\begin{proof}
    The nontrivial direction is
    right-to-left,
    so suppose that
    $\mathcal{T}\UniHyperDerivRel{\UniFLExtSCalc{\mathbf{w}}{\Sigma}} \UniSequentA$
    is
    witnessed by a deduction $\delta$.
    Obtain a standard deduction $\delta'$
    from $\delta$ by Lem.~\ref{lem:cut-restriction}.
    The new proof will
    have sequents over
    $\UniLangSet{\Sigma'}{\UniPropSet}$ only
    (by the subformula property proved also in Lem.~\ref{lem:cut-restriction}),
    and thus must be a deduction in 
    $\UniFLExtSCalc{\mathbf{w}}{\Sigma'}$.\qedhere
\end{proof}


\section{Hyper-Ackermannian lower bounds}
\label{sec:lower-bounds}
We obtain hyper-Ackermannian lower bounds
for the sequent-deducibility problem
in $\Sigma$-fragments of 
${\UniFLExtSCalc{\mathbf{w}}{}}$
where $\fus\in\Sigma$, via a reduction from the
reachability problem in lossy channel systems.
The result extends to the problem
of deducibility if
$\Sigma$ also contains $\ld$ (or $\rd$), $0$ and $1$, in view of Lem.~\ref{lem:der-seq-deducibility}.
We begin by introducing lossy channel systems
and their reachability problem,
and then show how to encode it
into sequent-deducibility
in ${\UniFLExtSCalc{\mathbf{w}}{\fus}}$.

\subsection{Lossy channel systems}
\label{sec:lossy-chnnel-systems}
A channel system is a computational model with unbounded FIFO channels (queues), defined as follows.

\begin{defn}
    A \emph{channel system} (CS)
    is a structure $
    \UniLcsA \UniSymbDef \UniStruct{\UniStateSet,\UniChannelSet,\UniMessAlphabet,\UniInstructionSet}$, where
    \begin{enumerate}
        \item $\UniStateSet \UniSymbDef \UniSet{q_1,\ldots,q_m}$ is a finite set of \emph{states};
        \item $\UniChannelSet \UniSymbDef \UniSet{c_1,\ldots,c_n}$
        is a finite set of \emph{channels};
        \item $\UniMessAlphabet \UniSymbDef \UniSet{\UniLetter{1},\ldots,\UniLetter{k}}$ is a finite \emph{message alphabet}; and
        \item $\UniInstructionSet \subseteq \UniStateSet \times \UniChannelSet \times \UniMessAlphabet \times \{ !, ? \} \times \UniStateSet$ is a finite set of \emph{instructions}.
    \end{enumerate}
We denote by $\UniConfSet{\UniLcsA}$ the set $\UniStateSet \times (\UniMessAlphabet^\ast)^n$ of \emph{configurations} of $\UniLcsA$,
where
$\UniMessAlphabet^\ast$
is the set of all finite sequences
of elements in $\UniMessAlphabet$.
\end{defn}


\begin{defn}
The \emph{perfect steps} of a CS $\UniLcsA$ are given by a binary relation $\UniPerfStep{\UniLcsA} \; \subseteq \UniConfSet{\UniLcsA} \times \UniConfSet{\UniLcsA}$ such that
$\UniTuple{q_i, u_1,\ldots, u_r ,\ldots,u_n} \UniPerfStep{\UniLcsA} \UniTuple{q_j, u_1,\ldots, v_r ,\ldots,u_n}$ if either
\begin{enumerate}
    \item $u_r =  \UniLetter{p}v_r$
    and $\UniTuple{q_i,c_r, \UniLetter{p}, ?,q_j} \in \UniInstructionSet$ (read and remove $\UniLetter{p}$ from the \emph{front} of channel $c_r$ and change state from $q_i$ to $q_j$)
    \item $v_r = u_r\UniLetter{p}$ and $\UniTuple{q_i,c_r, \UniLetter{p}, !,q_j} \in \UniInstructionSet$ (write $\UniLetter{p}$ at the \emph{back} of channel $c_r$ and change state from $q_i$ to $q_j$).
\end{enumerate}
Let $\UniPerfStep{\UniLcsA}^?$
and
$\UniPerfStep{\UniLcsA}^{!}$
denote the perfect steps given by items 1 and 2 respectively. Also let $\UniPerfStep{\UniLcsA}^\ast$ denote the reflexive
and transitive closure of $\UniPerfStep{\UniLcsA}$.
\end{defn}

Clearly, $\UniConfA \UniPerfStep{\UniLcsA}^\ast \UniConfB$
iff there is a finite sequence $\UniConfC_1,\ldots,\UniConfC_\eta$ ($\eta\geq 1$) of configurations  such that 
$u = \UniConfC_1$, $v = \UniConfC_\eta$
and $\UniConfC_i \UniPerfStep{\UniLcsA} \UniConfC_{i+1}$ for each~$i$ ($1 \leq i < \eta$).
This sequence is called a \emph{perfect computation in $\UniLcsA$} 
of length $\eta$ and it witnesses
that $\UniConfA$ \emph{reaches} $\UniConfB$ in $\UniLcsA$.

Reachability (as well as many other verification problems for CSs) is undecidable~\cite{brand1983}, but
adding lossy behaviour makes
it decidable.
This behaviour is introduced
in the operational semantics
of a CS as follows.

\begin{defn}
The \emph{lossy semantics} of a CS $\UniLcsA$
is given by a relation $\UniLossyStep{\UniLcsA} \; \subseteq \UniConfSet{\UniLcsA} \times \UniConfSet{\UniLcsA}$ 
that extends $\UniPerfStep{\UniLcsA}$
with all transitions 
(called \emph{lossy steps})
of the form
$\UniTuple{q_i, u_1,\ldots, u_r ,\ldots,u_n} \UniLossyStep{\UniLcsA}^l \UniTuple{q_i, u_1,\ldots, v_r,\ldots,u_n}$
such that
$u_r = w_1 \UniLetter{p} w_2$ and $v_r = w_1w_2$. Let $\UniLossyStep{\UniLcsA}^\ast$ be the reflexive
and transitive closure of $\UniLossyStep{\UniLcsA}$.
A channel system with a lossy semantics
is called a \emph{lossy channel system (LCS)}.
\end{defn}
Intuitively, 
in an LCS the
lossy steps permit any channel to lose a message (a symbol) from any position at any moment without changing the state. No instruction in the machine is required for this.
The relation $\UniLossyStep{\UniLcsA}^\ast$ induces a 
notion of \emph{lossy computation in $\UniLcsA$} and reachability is defined analogously to perfect computations
with $\UniPerfStep{\UniLcsA}^\ast$.
Since a step can be either read, write, or lossy, we use $\UniLossyStep{\UniLcsA}^{?}$, $\UniLossyStep{\UniLcsA}^!$
or $\UniLossyStep{\UniLcsA}^l$ to indicate precisely which step was performed
in a lossy computation.

\begin{defn}[Reachability problem in LCSs]
    Given a LCS $\UniLcsA$
    and configurations
    $\UniConfA,\UniConfB \in \UniConfSet{\UniLcsA}$,
    does $\UniConfA \UniLossyStep{\UniLcsA}^\ast \UniConfB$?
\end{defn}

\begin{remarkapx}
    \label{rem:lossy-equiv-std}
The relation $\UniLossyStep{\UniLcsA}^\ast$ has an equivalent definition in the literature~\cite[Sec. 2]{schnoebelen2002} as 
    $\UniConfA \rightsquigarrow_{\UniLcsA}^\ast \UniConfB$
    if, and only if,
    there are
    $\UniConfA',\UniConfB' \in \UniConfSet{\UniLcsA}$
    such that $\UniConfA \UniLossyStep{\UniLcsA}^{l\ast} \UniConfA'
    \UniPerfStep{\UniLcsA} \UniConfB'
    \UniLossyStep{\UniLcsA}^{l\ast} \UniConfB$ (the $\ast$ denotes the reflexive transitive closure).
    That is, messages may be lost before and after
    performing a perfect step.
    It can be shown that 
    $\UniLossyStep{\UniLcsA}^\ast \;=\;
    \UniLossyStep{\UniLcsA}^{l\ast} \circ \UniPerfStep{\UniLcsA} \circ \UniLossyStep{\UniLcsA}^{l\ast} \;=\;
    \rightsquigarrow_{\UniLcsA}^\ast$.
    The latter relation is shown~\cite[App. A]{chambart2008} to have no impact on the complexity of reachability compared to the ``write-lossy" mechanism employed in 
    the proof of Thm.~\ref{the:reachability-lcs-fww-complete}~\cite{chambart2008} stated below.
    
\end{remarkapx}


\begin{theoremapx}[{\cite[Thm. 5.5, Obs. 6.1, App. A]{chambart2008}}]
\label{the:reachability-lcs-fww-complete}
    Reachability in LCSs
    is
    $\UniHAck$-complete.
\end{theoremapx}

\subsection{Encoding reachability in LCSs into sequent-deducibility}
\label{sec:encoding}

We code each instruction of a lossy channel
system into a sequent. The collection of these sequents
is the \emph{theory} associated to
the system.

\begin{defn}
    \label{def:props-lcs}
    For an LCS
    $\UniLcsA =\UniStruct{\UniStateSet,\UniChannelSet,\UniMessAlphabet,\UniInstructionSet}$,
    define these sets of
    propositional variables:
    \begin{enumerate}
        \item $\UniStateVarSet \UniSymbDef
        \{ \UniStateVar{i} \mid
        q_i \in \UniStateSet
        \}$ (state variables)
        \item $\UniSepSet \UniSymbDef \{ \UniSepEnc{s}{i} \mid c_i\in\UniChannelSet\} \cup \{ \UniSepEnc{e}{i} \mid c_i \in \UniChannelSet\}$ (channel markers)
        \item $\UniChannelVarSet \UniSymbDef
        \{\UniChannelVar{a}
        \mid a \in \UniMessAlphabet\}$ (alphabet variables)
    \end{enumerate}
Moreover, define
$
\UniPropSetLcs{\UniLcsA}
\UniSymbDef
\UniStateVarSet
\cup 
\UniSepSet
\cup 
\UniChannelVarSet
$.
\end{defn}

\begin{defn}[Theory of an LCS]
\label{def:theory-lcs}
    Given an LCS 
    $\UniLcsA =\UniStruct{\UniStateSet,\UniChannelSet,\UniMessAlphabet,\UniInstructionSet}$,
    the \emph{theory of $\UniLcsA$},
    denoted by $\UniTheoryLCS{\UniLcsA}$,
    is defined as the union of the following finite sets
    of sequents:
    \begin{enumerate}
        \item 
        $\UniTheoryLCS{\UniLcsA}^!
        \UniSymbDef
        \{ \UniSequent{\UniSepEnc{e}{l},\UniStateVar{i}}{
        \UniChannelVar{a} \fus 
            (\UniSepEnc{e}{l} \fus
            \UniStateVar{j})
            }
        \mid \UniTuple{q_i, c_l, a, !, q_j} \in \UniInstructionSet \}$;
        \item 
        $\UniTheoryLCS{\UniLcsA}^?
        \UniSymbDef
        \{
        \UniSequent{\UniSepEnc{s}{l}, \UniChannelVar{a},\UniStateVar{i}}{
        \UniSepEnc{s}{l} \fus \UniStateVar{j} 
        }
        \mid \UniTuple{q_i, c_l, a, ?, q_j} \in \UniInstructionSet\}$;
        \item 
        $\UniTheoryLCS{\UniLcsA}^{\UniStateSet}
        \UniSymbDef
        \bigcup\limits_{\substack{Q \in \UniStateVarSet\\R \in \UniPropSetLcs{\UniLcsA}{\setminus}\UniStateVarSet}}
        \{ \UniSequent{R,\UniStateVar{}}{\UniStateVar{}\fus R} \} \cup
        \{ \UniSequent{\UniStateVar{},R}{R \fus \UniStateVar{}} \}
        $.
    \end{enumerate}
\end{defn}

\noindent Note that
$\UniTheoryLCS{\UniLcsA}$
is a regular theory in the sense of Def.~\ref{def:regular-theory-standard-cut}.

Recall that
$\bigotimes \UniList{\UniFmA_1,\ldots,\UniFmA_m} \UniSymbDef\UniFmA_1 \fus (\UniFmA_2 \fus ( \ldots ( \UniFmA_{m-1} \fus \UniFmA_m )\ldots))$, where $m \geq 1$
and $\UniFmA_1,\ldots,\UniFmA_m$
is a sequence of formulas.
We now translate a reachability instance
in a fixed LCS $\UniLcsA =\UniStruct{\UniStateSet,\UniChannelSet,\UniMessAlphabet,\UniInstructionSet}$
to a sequent-deducibility instance
in $\UniFLExtSCalc{\mathbf{w}}{\fus}$.
We begin by coding a configuration as a sequence.

\begin{defn}
    Given $\UniConfA \UniSymbDef \UniTuple{q_i, u_1,\ldots,u_n}
    \in \UniConfSet{\UniLcsA}$,
    define the sequence 
    $
        \UniEncConf{\UniConfA}
        \UniSymbDef
        \UniList{\UniStateVar{i},
        \UniSepEnc{s}{1},
        U_1,
        \UniSepEnc{e}{1},
        \ldots,
        \UniSepEnc{s}{n},
        U_n,
        \UniSepEnc{e}{n}}
    $, where, for $u_i = a_1 \cdots a_{k_i}\in \UniMessAlphabet^\ast$, $U_i$ is the sequence $\UniChannelVar{a_1}, \ldots, \UniChannelVar{a_{k_i}}$ of propositional variables.
\end{defn}


\begin{defn}
    For $\UniConfA \UniSymbDef \UniTuple{q_i, u_1,\ldots,u_n}$,  $
    \UniConfB \UniSymbDef \UniTuple{q_j, v_1,\ldots,v_n}
    \in \UniConfSet{\UniLcsA}$,
    define the sequent
    \[
        \UniEncProb{\UniConfB}{\UniConfA} 
        \UniSymbDef
        \UniSequent{\UniEncConf{\UniConfA}}{\bigotimes \UniEncConf{\UniConfB}}.
    \]
    Define also $\UniEncProbCom{\UniConfB}{\UniConfA}$
    as 
    the set of sequents of the above form
    where the occurrence of 
    $\UniStateVar{i}$ in the antecedent may appear in any position (i.e., not necessarily at the front as in $\UniEncConf{\UniConfA}$).
\end{defn}

We note that sequents in standard deductions in $\UniFLExtSCalc{\mathbf{w}}{\fus}$ from a regular theory have a nonempty succedent coming from the observation that every leaf has this property:

\begin{lemmaapxrep}\label{lem:empty-suc-der-cond}
    Let $\mathcal{T}$
    be a regular theory.
    It is never the case that
    $\mathcal{T}\UniHyperStdDerivRel{\UniFLExtSCalc{\mathbf{w}}{\fus}}  \UniSequent{\UniMSetFmA}{}$.
\end{lemmaapxrep}
\begin{proof}
    It follows because
    no axiomatic rule
    has conclusion with empty
    succedent
    and no other rule
    allows to derive a
    sequent with empty succedent
    from one with nonempty succedent.
\end{proof}

The \textit{flattening $\UniFlat{\UniFmA}$} is the sequence 
of propositional variables
obtained from a formula $\UniFmA$
by replacing `$\fus$' with `,'.
E.g., $\UniFlat{p \fus ((q \fus s) \fus r)} = p,q,s,r$.
The flattening $\UniFlat{\UniMSetFmA}$ of a sequence
$\UniMSetFmA$ of formulas is the concatenation of the flattening
of the formulas in $\UniMSetFmA$ in the order they appear in the sequence.
Let $\UniSepSet\UniMSetFmB$ denote the subsequence of $\UniMSetFmB$ having only variables from the set $\UniSepSet$
(that is, only channel markers).
The \emph{$\UniStateVarSet$-free subsequence} of a sequence of propositional variables
is obtained by deleting all
occurrences of state variables.

The following lemmas express that in standard deductions from a regular theory, in various situations, propositional variables occurring positively must also occur negatively.
They abstract what is needed for the right-to-left direction of Lem.~\ref{lem:encoding} below.
The proofs are by structural induction on standard deductions.

\begin{lemmaapx}
    \label{lem:state-free-sublist}
    Let $\UniMSetFmA$ be a sequence of 
    formulas, none having a state variable.
    If $\UniTheoryLCS{\UniLcsA}\UniHyperStdDerivRel{\UniFLExtSCalc{\mathbf{w}}{\fus}}  \UniSequent{\UniMSetFmA}{\UniFmA}$, then
    the $\UniStateVarSet$-free subsequence of
    $\UniFlat{\UniFmA}$ is a subsequence of 
    $\UniFlat{\UniMSetFmA}$.
\end{lemmaapx}
\begin{proof}
Structural induction on a standard deduction witnessing
$\UniTheoryLCS{\UniLcsA}\UniHyperDerivRel{\UniFLExtSCalc{\mathbf{w}}{\fus}} \UniSequent{\UniMSetFmA}{\UniFmA}$.
\emph{Base case}. For elements from
$\UniTheoryLCS{\UniLcsA}$ it holds vacuously since each sequent contains a state variable in the antecedent; for an identity sequent the claim is immediate.
\emph{Inductive step}. Suppose that the last rule instance applied is $\UniRule{\UniIRule}$, $\UniRuleSide{R}{\fus}$ or $\UniRuleSide{L}{\fus}$. Applying the IH to its premise(s), we have that the $\UniStateVarSet$-free subsequence of $\UniFlat{\UniFmA}$ is a subsequence of the flattening of the premise antecedent(s), and hence also of $\UniFlat{\UniMSetFmA}$. The last rule instance cannot be a standard cut as the antecedent of each element in 
$\UniTheoryLCS{\UniLcsA}$ contain a state variable hence $\UniMSetFmA$ would contains a state variable, contradicting the hypotheses.
The rule $\UniORule$
cannot occur due to Lem.~\ref{lem:empty-suc-der-cond}.
\end{proof}

\begin{lemmaapxrep}
    \label{lem:contiguous-seg-suc-to-ant}
    Let $\UniMSetFmA$ be a sequence of 
    formulas, $\UniConfA \in \UniConfSet{\UniLcsA}$
    and $L$ a nonempty subsequence of $\UniEncConf{\UniConfA}$.
    Then
    $\UniTheoryLCS{\UniLcsA}\UniHyperStdDerivRel{\UniFLExtSCalc{\mathbf{w}}{\fus}}  \UniSequent{\UniMSetFmA}{\bigotimes L}$
    implies that
    the variables in
    $\UniSepSet L$ 
    appear at least once in
    $\UniFlat{\UniMSetFmA}$.
\end{lemmaapxrep}
\begin{proof}
    Induction on the structure of a
    standard
    deduction witnessing 
    $\UniTheoryLCS{\UniLcsA}\UniHyperDerivRel{\UniFLExtSCalc{\mathbf{w}}{\fus}} \UniSequent{\UniMSetFmA}{\bigotimes L}$.
    \emph{Base case.}
     Straightforward: identity sequent $\UniSequent{p}{p}$ (trivial) and an element from $\UniTheoryLCS{\UniLcsA}$ (by inspection, it contains no channel variables or the same channel variable occurs in the antecedent and succedent).
    \emph{Inductive step}. Consider the last rule
    instance in the standard deduction.
For $\UniRuleSide{R}{\fus}$, $\UniRuleSide{L}{\fus}$ and $\UniRule{\UniIRule}$, we have by IH that each variable in $\UniSepSet L$ appears in $\UniFlat{\UniMSetFmA'}$ for the antecedent $\UniMSetFmA'$ of some premise of the rule instance.
Those variables will be carried to the conclusion antecedent. If the last rule applied is a standard cut, 
its right premise has the form $\UniSequent{\UniMSetFmA_1,\UniFmA,\UniMSetFmA_2}{\bigotimes L}$ where $\UniFmA$ is the cut formula. By IH, every variable in 
$\UniSepSet L$ appears in $\UniFlat{\UniMSetFmA_1,\UniFmA,\UniMSetFmA_2}$. 
Moreover, since the left premise is from  $\UniTheoryLCS{\UniLcsA}$, any channel variable occurring in the cut formula will occur in its antecedent, and hence also in $\UniFlat{\UniMSetFmA}$.
{The case of $\UniORule$
does not need to be considered in view of Lem.~\ref{lem:empty-suc-der-cond}}.
\end{proof}

\begin{lemmaapxrep}
    \label{lem:state-on-right-is-state-on-left}
        Let $\UniMSetFmA$ be a sequence of formulas
        and
        $\UniFmA$ be
        a formula.
    If $\UniTheoryLCS{\UniLcsA}\UniHyperStdDerivRel{\UniFLExtSCalc{\mathbf{w}}{\fus}}  \UniSequent{\UniMSetFmA}{Q_i \fus \UniFmA}$
    or
    $\UniTheoryLCS{\UniLcsA}\UniHyperStdDerivRel{\UniFLExtSCalc{\mathbf{w}}{\fus}}  \UniSequent{\UniMSetFmA}{Q_i}$, 
    then $\UniFlat{\UniMSetFmA}$ must contain
 a state variable.
\end{lemmaapxrep}
\begin{proof}
Structural induction on a
derivation witnessing 
$\UniTheoryLCS{\UniLcsA}\UniHyperDerivRel{\mathbf{SL_w^{\UniFusion}}} \UniSequent{\UniMSetFmA}{Q_i \fus \UniFmA}$.
\emph{Base case}. The case of an initial sequent is trivial, and the antecedent of any sequent from $\UniTheoryLCS{\UniLcsA}$ contains a state variable.
\emph{Inductive step}. Suppose that the last rule is $\UniRuleSide{R}{\fus}$. The left premise
has $Q_i$ as succedent, and hence by
IH, its antecedent will contain a state variable
that is carried down to the conclusion. The argument
is analogous for $\UniRule{\UniIRule}$ and $\UniRuleSide{L}{\fus}$.
If the last rule instance is a standard cut, its left premise is a sequent in $\UniTheoryLCS{\UniLcsA}$. As already observed, the antecedent of such a sequent contains a state variable and that will be carried down to the conclusion.
{The case of $\UniORule$
does not need to be considered in view of Lem.~\ref{lem:empty-suc-der-cond}}.

The case of
$\UniTheoryLCS{\UniLcsA}\UniHyperDerivRel{\UniFLExtSCalc{\mathbf{w}}{\fus}} \UniSequent{\UniMSetFmA}{Q_i}$ is similar.
\end{proof}

We are ready to reduce reachability 
in lossy channel systems
to sequent-deducibility.

\begin{lemmaapxrep}
    \label{lem:encoding}
    For all LCS $\UniLcsA$,
    given $\UniConfA,\UniConfB \in \UniConfSet{\UniLcsA}$,
    $
    \UniConfA \UniLossyStep{\UniLcsA}^\ast \UniConfB
    \text{ iff }
    \UniTheoryLCS{\UniLcsA}
    \UniHyperStdDerivRel{\UniFLExtSCalc{\mathbf{w}}{\fus}} 
    \UniSequentA
    $
    for some $\UniSequentA \in \UniEncProbCom{\UniConfB}{\UniConfA}$.
\end{lemmaapxrep}
\begin{inlineproof}
    In the left-to-right direction,
    we work by induction on 
    the length $k \geq 1$ of a computation
    witnessing
    $\UniConfA \UniLossyStep{\UniLcsA}^\ast \UniConfB$.
    \emph{Base case.}
    If $k=1$, $\UniConfA=\UniConfB$,
    and $\UniEncProb{\UniConfB}{\UniConfA}$ is
    provable by $\UniRule{(id)}$ and $\UniRuleSide{R}{\fus}$.
    \emph{Inductive step}. The computation decomposes as
    $\UniConfA \UniLossyStep{\UniLcsA}^\sigma \UniConfB' \UniLossyStep{\UniLcsA}^\ast \UniConfB$ with length $k+1$. 
    As
    $\UniConfB' \UniLossyStep{\UniLcsA}^\ast \UniConfB$
    is witnessed by a computation of length $k$, IH yields
    $\UniTheoryLCS{\UniLcsA}
\UniHyperDerivRel{\UniFLExtSCalc{\mathbf{w}}{\fus}}
\UniSequentA'$
    for some $\UniSequentA' \in \UniEncProbCom{\UniConfB}{\UniConfB'}$. We extend the latter deduction downwards to~$\UniSequentA$, using a case analysis on $\sigma$ to determine the form of $\UniSequentA'$ and $\UniSequentA$:
    
    \begin{enumerate}
        \item if $\sigma = \,?$, by effect of $\UniTuple{q_i, c_l, a, ?, q_j} \in \UniInstructionSet$:
        then 
        $u_l = a v'_l$,
        all the other channels
        have the same content,
        and 
        the state transits from $q_i$ to $q_j$.
        Thus $\UniSequentA'$ has the following form, possibly following repeated standard cuts with sequents in $\UniTheoryLCS{\UniLcsA}^{\UniStateSet}$.
        {
        \small
        \[
        \UniSequent{
        \UniSepEnc{s}{1},
        U_1,
        \UniSepEnc{e}{1},
        \ldots,
        \UniSepEnc{s}{l-1},
        U_{l-1},
        \UniSepEnc{e}{l-1},
            \clr{red}{
            \UniSepEnc{s}{l},
            \UniStateVar{j}
            },
        V'_l,
        \UniSepEnc{e}{l},
        \ldots,
        \UniSepEnc{s}{n},
        U_n,
        \UniSepEnc{e}{n}}{
        \bigotimes \UniEncConf{\UniConfB}}
        \]}
        Apply $\UniRuleSide{L}{\fus}$
        and a standard cut with
        $\UniSequent{
        \UniSepEnc{s}{l}, \UniChannelVar{a}, \UniStateVar{i}
        }{\clr{red}{\UniSepEnc{s}{l} \fus \UniStateVar{j}}}$ to obtain $\UniSequentA$.
        \item if $\sigma = l$:
        the state is unchanged, 
        and
        $v_l' = \clr{blue}{z_1z_2}$, for $u = \clr{blue}{z_1} \clr{red}{a} \clr{blue}{z_2}$.
        Thus $\UniSequentA'$ has the following form, possibly following repeated standard cuts with sequents in $\UniTheoryLCS{\UniLcsA}^{\UniStateSet}$.
        {\small
        \[
        \UniSequent{
        \UniStateVar{j},
        \UniSepEnc{s}{1},
        U_1,
        \UniSepEnc{e}{1},
        \ldots,
        \UniSepEnc{s}{l},
            \clr{blue}{
                Z_1,
                Z_2
            },
        \UniSepEnc{e}{l},
        \ldots,
        \UniSepEnc{s}{n},
        U_n,
        \UniSepEnc{e}{n}
        }{
        \bigotimes \UniEncConf{\UniConfB}}
        \]}
        By $\UniRule{\UniIRule}$
        $
        \UniSequent{
        \UniStateVar{j},
        \UniSepEnc{s}{1},
        U_1,
        \UniSepEnc{e}{1},
        \ldots,
        \UniSepEnc{s}{l},
            \clr{blue}{
                Z_1,
                \clr{red}{\UniChannelVar{a}},
                Z_2
            },
        \UniSepEnc{e}{l},
        \ldots,
        \UniSepEnc{s}{n},
        U_n,
        \UniSepEnc{e}{n}
        }{
        \bigotimes \UniEncConf{\UniConfB}}
        $
        \item
        if $\sigma = \, !$, by effect of $\UniTuple{q_i, c_l, a, !, q_j} \in \UniInstructionSet$:
        by definition of this step, $v'_l = u_la$,
        the content in other channels of $\UniConfA$ and $\UniConfB'$ is unchanged,
        and the state transits from $q_i$ to $q_j$.
        Thus $\UniSequentA'$ has the following form modulo the position of $\UniStateVar{j}$. In any case, the $\UniSequentA'$ can be moved to the displayed position through repeated standard cuts with sequents in $\UniTheoryLCS{\UniLcsA}^{\UniStateSet}$.
        \[
        \UniSequent{
        \UniSepEnc{s}{1},
        U_1,
        \UniSepEnc{e}{1},
        \ldots,
        \UniSepEnc{s}{l},
            U_l,
            \clr{red}{
        \UniChannelVar{a},
            \UniSepEnc{e}{l}, \UniStateVar{j}},
            \UniSepEnc{s}{l+1},
        \ldots,
        \UniSepEnc{s}{n},
        U_n,
        \UniSepEnc{e}{n}}{
        \bigotimes \UniEncConf{\UniConfB}}
        \]
        It suffices now to apply $\UniRuleSide{L}{\fus}$ twice and a standard cut with
        $\UniSequent{\UniSepEnc{e}{l},\UniStateVar{i}}{\clr{red}{
        \UniChannelVar{a} \fus 
            \UniSepEnc{e}{l} \fus \UniStateVar{j}}}$ to obtain the sequent $\UniSequentA$: 
$\UniSequent{
        \UniSepEnc{s}{1},
        U_1,
        \UniSepEnc{e}{1},
        \ldots,
        \UniSepEnc{s}{l},
            U_l,
            {
            \UniSepEnc{e}{l}, \UniStateVar{j}},
            \UniSepEnc{s}{l+1},
        \ldots,
        \UniSepEnc{s}{n},
        U_n,
        \UniSepEnc{e}{n}}{
        \bigotimes \UniEncConf{\UniConfB}}$.
    \end{enumerate}

    For the right-to-left
    direction,
    we establish property
    $\mathcal{P}(\delta)
    \UniSymbDef$:
    ``for all $\UniConfA, \UniConfB
    \in \UniConfSet{\UniLcsA}$,
    if
    $\delta$ 
    is a standard deduction
    witnessing
    $\UniTheoryLCS{\UniLcsA}
\UniHyperDerivRel{\UniFLExtSCalc{\mathbf{w}}{\fus}}
    \UniSequentA$
    for some $\UniSequentA \in \UniEncProbCom{\UniConfB}{\UniConfA}$,
    then $\UniConfA \UniLossyStep{\UniLcsA}^\ast \UniConfB$''.
    Induction on the structure of $\delta$.
    
\emph{Base case}. No initial sequent
    nor any sequent from
    $\UniTheoryLCS{\UniLcsA}$
    is an image of an encoding
    of a configuration since it does not contain any pair $\{\UniSepEnc{s}{i},\UniSepEnc{e}{i}\}$ of channel markers.

    \emph{Inductive step}. 
    Let
    $\UniConfA = (q_i, u_1,\ldots,u_n)$
    and
    $\UniConfB = (q_j, v_1,\ldots,v_n)$.
    By cases on the last rule applied in $\delta$.
    It cannot be $\UniRuleSide{L}{\fus}$ since the antecedent of $\UniSequentA$ is a sequence of propositional variables.
    Then it could be:

    \begin{enumerate}
        \item $\UniORule$:
        Not possible by
        Lem.~\ref{lem:empty-suc-der-cond} since $\delta$ is a standard deduction.
        \item $\UniRuleSide{R}{\fus}$:
        Write $\UniSequentA$ as $\UniSequent{\UniMSetFmA_1,\UniMSetFmA_2}{Q_j \fus \UniFmA_2}$. The premises then are $\UniSequent{\UniMSetFmA_1}{Q_j}$ and $\UniSequent{\UniMSetFmA_2}{\UniFmA_2}$.         
By Lem.~\ref{lem:state-on-right-is-state-on-left},
        $\UniMSetFmA_1$ must have a formula
        with a state variable,
        and thus $\UniMSetFmA_2$
        lacks a state variable.
        If $\UniMSetFmA_1$ contained anything more, it must contain a channel marker.
        By Lem.~\ref{lem:state-free-sublist},
        $\UniMSetFmA_2$ must contain all the
        channel markers to fulfil the requirement
        that $\UniFlat{\UniFmA_2}$ must
        be a subsequence of $\UniFlat{\UniMSetFmA_2}$.
        Hence $\UniMSetFmA_1$ is the singleton sequence $Q_l$.
        Now, for $\UniSequent{\UniStateVar{l}}{\UniStateVar{j}}$ to be provable,
        we must have $l = j$.
        Also, $\UniFlat{\UniFmA_2}$ a subsequence of $\UniFlat{\UniMSetFmA_2}$
        implies that the
        channel portions in the succedent
        are subsequences of the corresponding ones
        in the antecedent. Thus, $\UniConfA$ can reach $\UniConfB$
        by losing messages.
        Observe: the IH is not employed.

        \item $\UniIRule$: 
        We claim that the weakening introduces
        a variable into the channel portion.
        This suffices since 
        the premise would then be the image of
        a configuration that, by IH, reaches $\UniConfB$,
        and is itself reachable from $\UniConfA$
        by a lossy step. To establish the claim, first note that the sequents involved in
        an application
        of $\UniIRule$ have the same
        succedent, in this case
        $\bigotimes \UniEncConf{\UniConfB}$.
        Let $\UniMSetFmA'$ be the premise antecedent.
        By Lem.~\ref{lem:contiguous-seg-suc-to-ant},
        we have that all channel variables
        appear in $\UniMSetFmA'$.
        Since these symbols appear only once
        in conclusion antecedent, weakening could not
        have introduced them.
        If the weakening introduced the state variable, then $\UniMSetFmA'$ would
        have no state variables; that is impossible due to Lem.~\ref{lem:state-on-right-is-state-on-left} since the succedent
        does contain a state variable.
        The remaining possibility is that weakening introduced a variable in the channel portion, so the claim is proved. 
        \item Standard cut with a sequent in $\UniTheoryLCS{\UniLcsA}^{?}$: The deduction has the form
        \begin{center}
            \begin{footnotesize}
            \AxiomC{}
            \RightLabel{$\UniTheoryLCS{\UniLcsA}^{?}$}
            \UnaryInfC{$\UniSequent{\clr{blue}{ \UniSepEnc{s}{l}, \UniChannelVar{a},\UniStateVar{i}}}{\clr{red}{{\UniSepEnc{s}{l} \fus \UniStateVar{p}}}}$}
            \AxiomC{$\delta_1$}
            \noLine
            \UnaryInfC{
$\UniSequent{
        \UniSepEnc{s}{1},
        U_1,
        \UniSepEnc{e}{1},
        \ldots,
        \clr{red}{
        \UniSepEnc{s}{l}\fus
        Q_p
        },
        U_l,
        \UniSepEnc{e}{l},
        \ldots,
        \UniSepEnc{s}{n},
        U_n,
        \UniSepEnc{e}{n}}{
        \bigotimes \UniEncConf{\UniConfB}}$
            }
            \RightLabel{$\UniRule{(cut)}$}
            \BinaryInfC{$\UniSequent{
        \UniSepEnc{s}{1},
        U_1,
        \UniSepEnc{e}{1},
        \ldots,
        \clr{blue}{
        \UniSepEnc{s}{l},
        \UniChannelVar{a},
        \UniStateVar{i}
        },
        U_l,
        \UniSepEnc{e}{l},
        \ldots,
        \UniSepEnc{s}{n},
        U_n,
        \UniSepEnc{e}{n}}{
        \bigotimes \UniEncConf{\UniConfB}}$}
            \DisplayProof
            \end{footnotesize}
        \end{center}
        Consider the last rule in $\delta_1$. It cannot be a (standard) cut as that would have introduced a state variable on its own (i.e., unfused) into the conclusion. If it was
        $\UniRuleSide{R}{\fus}$ then 
        the left premise has the form
        $\UniSequent{\UniMSetFmA_1}{Q_j}$
        and the second premise has the form
        $\UniSequent{\UniMSetFmA_2}{\UniFmA_2}$.
        By Lem.~\ref{lem:state-on-right-is-state-on-left},
        $\clr{red}{
            \UniSepEnc{s}{l} \fus
            \UniStateVar{p}
            } \in \UniMSetFmA_1$,
            and thus $\UniMSetFmA_2$
            consists only of
            propositional variables.
            Since $\UniSepEnc{s}{l}$ is not in $\UniMSetFmA_2$ but $\UniFmA_2$ contains all channel markers, 
            Lem.~\ref{lem:contiguous-seg-suc-to-ant} gives a contradiction. So $\UniRuleSide{R}{\fus}$ was not the last rule applied. Neither can it be $\UniORule$
        due to Lem.~\ref{lem:empty-suc-der-cond}.

        We are left with two alternatives for the last rule in $\delta_1$:
        $\UniIRule$ and $\UniRuleSide{L}{\fus}$.
If $\UniIRule$, arguing as in the previous item, the introduced variable must be in the channel portion by Lem.~\ref{lem:contiguous-seg-suc-to-ant}
        and Lem.~\ref{lem:state-on-right-is-state-on-left}.
        Iterating the entire argument up to this point (implicit induction), after multiple applications of $\UniIRule$ we must ultimately encounter $\UniRuleSide{L}{\fus}$ (we have ruled out all the other possibilities). The situation is therefore the following.
%
                \begin{center}
            \begin{footnotesize}
            \AxiomC{}
            \RightLabel{$\UniTheoryLCS{\UniLcsA}^{?}$}
            \UnaryInfC{$\UniSequent{\clr{blue}{
            \UniSepEnc{s}{l}, 
            \UniChannelVar{a},
            \UniStateVar{i}
            }
            }{\clr{red}{
            {
             \UniSepEnc{s}{l}\fus
             \UniStateVar{p}
            }}}$}
            \AxiomC{$\delta_1'$}
            \noLine
            \UnaryInfC{$\UniSequent{
        \UniSepEnc{s}{1},
        U'_1,
        \UniSepEnc{e}{1},
        \ldots,
        \clr{red}{
        \UniSepEnc{s}{l},
        \UniStateVar{p}
        },
        U'_l,
        \UniSepEnc{e}{l},
        \ldots,
        \UniSepEnc{s}{n},
        U'_n,
        \UniSepEnc{e}{n}
        }{
        \bigotimes \UniEncConf{\UniConfB}}$}
        \RightLabel{$\UniRuleSide{L}{\fus}$}
        \UnaryInfC{$\UniSequent{
        \UniSepEnc{s}{1},
        U'_1,
        \UniSepEnc{e}{1},
        \ldots,
        \clr{red}{{
        \UniSepEnc{s}{l} \fus \UniStateVar{p} 
        }},
        U'_l,
        \UniSepEnc{e}{l},
        \ldots,
        \UniSepEnc{s}{n},
        U'_n,
        \UniSepEnc{e}{n}
        }{
        \bigotimes \UniEncConf{\UniConfB}}$}
        \RightLabel{$\UniIRule^\ast$}
        \doubleLine
            \UnaryInfC{
$\UniSequent{
        \UniSepEnc{s}{1},
        U_1,
        \UniSepEnc{e}{1},
        \ldots,
        \clr{red}{{
        \UniSepEnc{s}{l} \fus \UniStateVar{p} 
        }},
        U_l,
        \UniSepEnc{e}{l},
        \ldots,
        \UniSepEnc{s}{n},
        U_n,
        \UniSepEnc{e}{n}
        }{
        \bigotimes \UniEncConf{\UniConfB}}$
            }
            \RightLabel{$\UniRule{(cut)}$}
            \BinaryInfC{$\UniSequent{
        \UniSepEnc{s}{1},
        U_1,
        \UniSepEnc{e}{1},
        \ldots,
        \clr{blue}{
        \UniSepEnc{s}{l},
        \UniChannelVar{a},
        \UniStateVar{i}
        },
        U_l,
        \UniSepEnc{e}{l},
        \ldots,
        \UniSepEnc{s}{n},
        U_n,
        \UniSepEnc{e}{n}
        }{
        \bigotimes \UniEncConf{\UniConfB}}$}
            \DisplayProof
            \end{footnotesize}
        \end{center}
        
        From the IH, we obtain that
        $\UniTuple{q_p, u'_1, \ldots, u'_l, \ldots, u'_n} \UniLossyStep{\UniLcsA}^\ast \UniConfB$. Since
        $\UniTuple{q_i, u_1, \ldots, a u_l, \ldots, u_n}\UniLossyStep{\UniLcsA}^{?} \UniTuple{q_p, u_1, \ldots, u_l, \ldots, u_n}\UniLossyStep{\UniLcsA}^\ast \UniTuple{q_p, u'_1, \ldots, u'_l, \ldots, u'_n}$,
        we are done.
                    \item Standard cut with a sequent in $\UniTheoryLCS{\UniLcsA}^{!}$: In this case
        the deduction has the form
        \begin{center}
            \footnotesize
            \AxiomC{}
            \RightLabel{$\UniTheoryLCS{\UniLcsA}^{!}$}
            \UnaryInfC{$\UniSequent{
                \clr{blue}{
                \UniSepEnc{e}{l},
                \UniStateVar{i}
                }
                }{{
            \clr{red}{
        \UniChannelVar{a} \fus 
            \UniSepEnc{e}{l} \fus
            \UniStateVar{p}
            }}}$}
            \AxiomC{$\delta_1$}
            \noLine
            \UnaryInfC{
$\UniSequent{
        \UniSepEnc{s}{1},
        U_1,
        \UniSepEnc{e}{1},
        \ldots,
        \UniSepEnc{s}{l},
        U_l,
        \clr{red}{
        \UniChannelVar{a} \fus
        \UniSepEnc{e}{l} \fus
        \UniStateVar{p}
        },
        \ldots,
        \UniSepEnc{s}{n},
        U_n,
        \UniSepEnc{e}{n}}{
        \bigotimes \UniEncConf{\UniConfB}}$
            }
            \RightLabel{$\UniRule{(cut)}$}
            \BinaryInfC{$\UniSequent{
        \UniSepEnc{s}{1},
        U_1,
        \UniSepEnc{e}{1},
        \ldots,
        \UniSepEnc{s}{l},
        U_l,
        \clr{blue}{
        \UniSepEnc{e}{l},
        \UniStateVar{i}
        },
        \ldots,
        \UniSepEnc{s}{n},
        U_n,
        \UniSepEnc{e}{n}}{
        \bigotimes \UniEncConf{\UniConfB}}$}
            \DisplayProof
        \end{center}
            
        Consider the last rules in $\delta_1$.
        Arguing as in the $\UniTheoryLCS{\UniLcsA}^{?}$ case, it is a sequence of $\UniIRule$ rules introducing variables in channel portions, then
        $\UniRuleSide{L}{\fus}$.
        The latter premise is
        \[\UniSequent{
        \UniSepEnc{s}{1},
        U'_1,
        \UniSepEnc{e}{1},
        \ldots,
        \UniSepEnc{s}{l},
        U'_l,
        \clr{red}{
        \UniChannelVar{a},
        \UniSepEnc{e}{l} \fus
        \UniStateVar{p}
        },
        \ldots,
        \UniSepEnc{s}{n},
        U'_n,
        \UniSepEnc{e}{n}}{
        \bigotimes \UniEncConf{\UniConfB}}\]

        Since no state variable occurs
        by itself in the antecedent, the above sequent was not derived by a (standard) cut.
        Nor was it by $\UniRuleSide{R}{\fus}$,
        since its left premise would contain $\UniSepEnc{e}{l} \fus \UniStateVar{p}$ hence depriving the right premise antecedent of $\UniSepEnc{e}{l}$ leading to a contradiction with Lem.~\ref{lem:contiguous-seg-suc-to-ant}.
        Nor was it $\UniORule$ due to Lem.~\ref{lem:empty-suc-der-cond}.
        Hence we encounter once again a sequence of
        $\UniIRule$ rules followed by an application of
        $\UniRuleSide{L}{\fus}$, resulting in the sequent

        \[\UniSequent{
        \UniSepEnc{s}{1},
        U''_1,
        \UniSepEnc{e}{1},
        \ldots,
        \UniSepEnc{s}{l},
        U''_l,
        \clr{red}{
        \UniChannelVar{a},
        \UniSepEnc{e}{l},
        \UniStateVar{p}
        },
        \ldots,
        \UniSepEnc{s}{n},
        U''_n,
        \UniSepEnc{e}{n}}{
        \bigotimes \UniEncConf{\UniConfB}}\]

By IH, we obtain that
        $\UniTuple{q_p, u''_1,\ldots,u''_la,\ldots,u''_n} \UniLossyStep{\UniLcsA}^\ast\UniConfB$.
Since $\UniTuple{q_i, u_1,\ldots,u_l,\ldots,u_n} \UniLossyStep{\UniLcsA}^{!}\UniTuple{q_p, u_1,\ldots,u_la,\ldots,u_n} \UniLossyStep{\UniLcsA}^\ast\UniTuple{q_p, u''_1,\ldots,u''_la,\ldots,u''_n}$, 
we conclude
        $\UniConfA \UniLossyStep{\UniLcsA}^\ast \UniConfB$.

                \item Standard cut with $\UniSequent{\UniStateVar{i},R}{R \fus \UniStateVar{i}}\in\UniTheoryLCS{\UniLcsA}^{\UniStateSet}$
        (sequent $\UniSequent{R,\UniStateVar{i}}{\UniStateVar{i}\fus R}$ is similar):

                \begin{center}
                \begin{footnotesize}
            \AxiomC{}
            \RightLabel{$\UniTheoryLCS{\UniLcsA}^{\UniStateSet}$}
            \UnaryInfC{$\UniSequent{\clr{blue}{\UniStateVar{i},R}}{\clr{red}{R \fus \UniStateVar{i}}}$}
            \AxiomC{$\delta_1$}
            \noLine
            \UnaryInfC{$\UniSequent{\UniMSetFmA_1,\clr{red}{R \fus \UniStateVar{i}},\UniMSetFmA_2}{\bigotimes \UniEncConf{\UniConfB}}$}
            \RightLabel{$\UniRule{(cut)}$}
            \BinaryInfC{$\UniSequent{\UniMSetFmA_1, \clr{blue}{\UniStateVar{i},R},\UniMSetFmA_2}{\bigotimes \UniEncConf{\UniConfB}}$}
            \DisplayProof
        \end{footnotesize}
        \end{center}

         As in the previous cases,
        the last rule applied in $\delta_1$ could not have been a
        standard cut nor $\UniORule$.
        Suppose it is $\UniRuleSide{R}{\fus}$. Writing the endsequent of $\delta_1$ as
        $\UniSequent{\UniMSetFmA_3,\UniMSetFmA_4}{Q_j \fus \UniFmA_2}$, the left premise would have the form $\UniSequent{\UniMSetFmA_3}{Q_i}$ where $\UniMSetFmA_3$ must contain $R \fus \UniStateVar{i}$ (Lem.~\ref{lem:state-on-right-is-state-on-left}) and hence also $\UniMSetFmA_1$. If $\UniMSetFmA_1$ is non-empty then it must contain a channel marker and hence in the right premise $\UniSequent{\UniMSetFmA_4}{\UniFmA_2}$, the $\UniMSetFmA_4$ would lack a channel marker contradicting Lem.~\ref{lem:contiguous-seg-suc-to-ant}. If $\UniMSetFmA_1$ is empty then $R$ must be a channel marker and once again $\UniMSetFmA_4$ would lack a channel marker leading to a contradiction. Therefore the rule is not $\UniRuleSide{R}{\fus}$. 
        
        The remaining possibility is a sequence of $\UniIRule$ followed by $\UniRuleSide{L}{\fus}$.
        The premise of the latter is in the image of an encoding, thus
        the IH applies and we are done.\qedhere

    \end{enumerate}
\end{inlineproof}

\begin{theoremapxrep}\label{thm:lbs}
    Let $\Sigma \subseteq \UniSigFL{}$.
If $\fus \in \Sigma$, then sequent-deducibility in
$\UniFLExtSCalc{\mathbf{w}}{\Sigma}$ is 
$\UniHAck$-hard. Moreover, if $\{ \fus, \ld, 0, 1 \} \subseteq \Sigma$, then deducibility is $\UniHAck$-hard.
    
\end{theoremapxrep}
\begin{proof}
    The first assertion
    follows from 
    Lem.~\ref{lem:encoding}, plus
    the fact that the reduction established there is polynomial-time,
    and Lem.~\ref{lem:subsig-sig-der-seq-equiv}.
    The second one follows from
    the first
    and Lem.~\ref{lem:der-seq-deducibility}.
%
%
\end{proof}

\section{Hyper-Ackermannian upper bounds}
\label{sec:upper-bounds}
We give an algorithm
to decide the relation $\UniHyperDerivRel{\UniSeqCalcA}$
for regular theories,
for a sequent calculus $\UniSeqCalcA$
 containing the rule $\UniIRule$
and satisfying a generalized subformula property as defined below.
In what follows,
given a finite $\UniSubfmlaHyperseqSet\subseteq\UniLangSet{\Sigma}{\UniPropSet}$,
an \emph{$\UniSubfmlaHyperseqSet$-sequent}
is a sequent in which only formulas in $\UniSubfmlaHyperseqSet$
appear.

\begin{defn}
    \label{def:amenable-seq-calc}
    A sequent calculus that is a structural rule extension of $\UniFLExtSCalc{\UniWProp}{\Sigma}$
($\Sigma \subseteq \UniSigFL{}$) is called 
\emph{amenable}
    provided it satisfies
    the \emph{generalized subformula property}:
    for every regular theory
    $\mathcal{T}$, it is the case that
    $\mathcal{T} \UniHyperDerivRel{\UniSeqCalcA} \UniSequentA$
    iff there
    is a $\mathcal{T}$-deduction of $\UniSequentA$
    where only
    $\UniSubf{\mathcal{T} \cup \{ \UniSequentA \} }$-sequents
    appear
    (such a deduction is called \emph{analytic}).
\end{defn}
$\UniFLExtSCalc{\UniWProp}{\Sigma}$
is amenable for any 
$\Sigma \subseteq \UniSigFL{}$;
Lem.~\ref{lem:cut-restriction} gives the generalized subformula property. Indeed, any \emph{analytic structural rule} extension of an amenable calculus is amenable (see results for the substructural hierarchy, Ciabattoni et al.~\cite{CiaGalTer17}).



\subsection{Just enough on well-quasi-order theory}

A \emph{quasi-ordered set (qo-set)} 
	is a structure $\UniWqoA\UniSymbDef\UniStruct{\UniWqoSet, \UniWqoRel{\UniWqoA}}$,
	where
		$\UniWqoSet$ is a set
		 and $\UniWqoRel{\UniWqoA} \, \subseteq \UniWqoSet \times \UniWqoSet$ is reflexive and transitive (a \emph{quasi order}, for short).
   Abusing notation, we write $\UniValA \in \UniWqoA$
   for $\UniValA \in \UniWqoSet$.
   We denote by
   $\UniNaturalSet$ the
   set of natural numbers;
   by
   $\UniNaturalSetNN$
   the set $\UniNaturalSet{\setminus}\{0\}$;
   and, given
   $\ell \in \UniNaturalSet \cup \{ \omega\}$, we write
$\UniNaturalInitSeg{\UniLengthSequence}$ for the set
$\{ \UniValA \in \UniNaturalSet \mid \UniValA < \ell\}$.
	Given a qo-set $\UniWqoA$,
 a \emph{bad sequence over $\UniWqoA$} (of length $\UniLengthSequence$)
	is a sequence $(\UniValA_i)_{i \in \UniNaturalInitSeg{\UniLengthSequence}}$ of elements of $\UniWqoA$
	such that, for all $i < j$,
 $a_i \not\UniWqoRel{\UniWqoA} a_j$.
%
	A qo-set $\UniWqoA$ is a \emph{well-quasi-ordered set (wqo-set)}
	if every bad sequence over it is finite~(see~\cite[Sec. 2]{higman1952} for equivalent definitions).
 Examples include 
 $\mathbb{N}^k$ under component-wise ordering (Dickson's Lemma~\cite{dickson1913}), and
 sequences under the subword embedding ordering:
 

 \begin{theoremapx}[Higman's Lemma~\cite{higman1952}]\label{thm:higman-lemma}
     For any finite set $\UniSubfmlaHyperseqSet$,
     $\UniStruct{\UniSubfmlaHyperseqSet^\ast,\UniWqoRel{\ast}}$
     is a wqo, where, recall, 
     $\UniSubfmlaHyperseqSet^\ast$
     is the set of all finite sequences
     of elements from $\UniSubfmlaHyperseqSet$
     and
     $w_1 \UniWqoRel{\ast} w_2$ 
    iff
    $w_1$ is obtainable from $w_2$ by deleting some
    elements.
 \end{theoremapx}

Bad sequences over a wqo, though finite, do not in general have a maximum length. For example, $(1,0),(0,k),(0,k-1),\ldots,(0,0)$ is a bad sequence on $\mathbb{N}^2$ under the component-wise ordering with length $k+2$ for every $k\in\mathbb{N}$. Nevertheless, a maximum length can be ensured by limiting the size of each element in the sequence to some (fixed) function of the size of the preceding element. To achieve this, the wqo is enriched with more structure.
A \emph{normed \mbox{(well-)quasi-ordered}
 set (n(w)qo-set)}~\cite{schmitz2012notes}
	is a structure $\UniWqoA \UniSymbDef \UniStruct{\UniWqoSet, \UniWqoRel{\UniWqoA}, \UniNorm{\cdot}{\UniWqoA}}$,
	where
		$\UniStruct{\UniWqoSet, \UniWqoRel{\UniWqoA}}$ is a (w)qo-set and
		 $\UniNorm{\cdot}{\UniWqoA} : \UniWqoSet \to \UniNaturalSet$ is a \emph{proper norm},
   meaning that, for all $n \in \UniNaturalSet$,
			$
   \UniSet{\UniValA \in \UniWqoSet : \UniNorm{\UniValA}{\UniWqoA} < n}$ is finite.
   It is easy to check that the following is a nwqo.

   \begin{defn}\label{def:hig-nwqo}
            Given a finite set $\UniSubfmlaHyperseqSet$,
     let $\UniHigFinAlph{\UniSubfmlaHyperseqSet}
     \UniSymbDef \UniStruct{\UniSubfmlaHyperseqSet^\ast, \UniWqoRel{\ast}, \UniNorm{\cdot}{\ast}}$,
     where $\UniNorm{w}{\ast}$ is
     the length of $w$
     (and $\UniStruct{\UniSubfmlaHyperseqSet^\ast, \UniWqoRel{\ast}}$ is as in Thm.~\ref{thm:higman-lemma}).
   \end{defn}

   Indeed, the
   \emph{disjoint sum} of these nwqos, defined below, is also an nwqo~\cite{schmitz2012notes}.

   \begin{defn}
       Given a finite set $\UniSubfmlaHyperseqSet$
       and $m \in \UniNaturalSet$,
       let $m \cdot \UniHigFinAlph{\UniSubfmlaHyperseqSet}
       \UniSymbDef \UniStruct{
       m \cdot \UniSubfmlaHyperseqSet^\ast, 
       \UniWqoRel{m \cdot \UniHigFinAlph{\UniSubfmlaHyperseqSet}}, \UniNorm{\cdot}{m \cdot \UniHigFinAlph{\UniSubfmlaHyperseqSet}}}$.
       Here, $m \cdot \UniSubfmlaHyperseqSet^\ast \UniSymbDef \{ 1, \ldots, m \} \times \UniSubfmlaHyperseqSet^\ast$
       is the disjoint sum of $m$ copies of $\UniSubfmlaHyperseqSet^\ast$, and
       $(i, w_1) \UniWqoRel{m \cdot \UniHigFinAlph{\UniSubfmlaHyperseqSet}} (j, w_2)$
       iff $i = j$ and $w_1 \UniWqoRel{\ast} w_2$. Also, define
       $\UniNorm{(i, w)}{m \cdot \UniHigFinAlph{\UniSubfmlaHyperseqSet}} \UniSymbDef \UniNorm{w}{\ast}$.
   \end{defn}

	A \emph{control function} is a mapping $\UniControlFunctionA : \UniNaturalSet \to \UniNaturalSet$
	that is strictly increasing (for all $\UniValA,\UniValB \in \UniNaturalSet$, if $\UniValA < \UniValB$, then $\UniControlFunctionA(\UniValA) < \UniControlFunctionA(\UniValB)$)
    and strictly inflationary
	($\UniValA < \UniControlFunctionA(\UniValA)$ for all $\UniValA \in \UniNaturalSet$).

	For an nwqo-set $\UniWqoA$,
control function $\UniControlFunctionA$,
 and \emph{initial parameter} $\UniControlParam \in \UniNaturalSet$,
	a \emph{$(\UniControlFunctionA, \UniControlParam)$-controlled bad sequence over $\UniWqoA$} 
	is a bad sequence $\UniValA_0,\ldots,\UniValA_{\UniLengthSequence-1}$ over $\UniWqoA$ where
	$\UniNorm{\UniValA_i}{\UniWqoA} < \UniControlFunctionA^i(\UniControlParam)$ for all $0 \leq i < \UniLengthSequence$.
The $(\UniControlFunctionA, \UniControlParam)$-controlled bad sequences have a maximum length~\cite{figueira2011}
using K\"onig's Lemma.
Denote by $\UniLengFunc{\UniWqoA}{\UniControlFunctionA}(t)$
 the maximum length
 of a $(\UniControlFunctionA,\UniControlParam)$-controlled
 bad sequence over $\UniWqoA$.
 We call
 $\UniLengFunc{\UniWqoA}{\UniControlFunctionA}
 : \UniNaturalSet \to \UniNaturalSet$
 the \emph{length function}
 of $\UniWqoA$ (for $\UniControlFunctionA$).
Bounds for $(\UniControlFunctionA, \UniControlParam)$-controlled bad sequences over $\UniWqoA$
reduce to finding bounds for $\UniLengFunc{\UniWqoA}{\UniControlFunctionA}$ (`\emph{length theorems}'),
expressed in terms of the ordinal-indexed \emph{extended Grzegorczyk hierarchy} 
$\{ \UniGrzHLevel{\alpha}  \}_{\alpha}$, from which
the hierarchy $\{ \UniFGHProbOneAppLevel{\alpha} \}_\alpha$
of \emph{fast-growing complexity classes} are defined
(the reader is referred to~\cite[Sec.~2]{schmitz2016hierar} for detailed definitions).
What matters here is that to classify a problem in $\UniFGHProbOneAppLevel{\alpha}$, it suffices to show that the complexity of every instance is upper bounded by a function in $\UniGrzHLevel{\beta}$ for some $\beta < \alpha$.
Schmitz and Schnoebelen obtained
the following length theorem for Higman's lemma.

\begin{theoremapx}[{\cite[Thm.~5.3]{schmitz2011}}]\label{thm:hig-leng-the}
    For any finite set $\UniSubfmlaHyperseqSet$, $m \in \UniNaturalSet$,
    and primitive recursive control function
    $\UniControlFunctionA$, the length function
    $\UniLengFunc{m \cdot \UniHigFinAlph{\UniSubfmlaHyperseqSet}}{\UniControlFunctionA}$
    is upper bounded by a function 
    in 
    $\UniGrzHLevel{\omega^{\UniSetCard{\UniSubfmlaHyperseqSet}-1}}$.
\end{theoremapx}

A \textit{reflection} is a map between nqo-sets that preserves bad sequences. Hence, its existence implies the reverse transfer of the wqo-property and upper bounding of length theorem.
In the next subsection, we obtain a length theorem for a nwqo over sequents in this way.

\begin{defn}[Reflection]
\label{def:reflection}
	Let $\UniWqoA_1 \UniSymbDef \UniStruct{\UniWqoSet_1, \UniWqoRel{1}, \UniNorm{\cdot}{1}}$ 
	and $\UniWqoA_2 \UniSymbDef \UniStruct{\UniWqoSet_2, \UniWqoRel{2},\UniNorm{\cdot}{2}}$ be nqo-sets.
	A \emph{reflection} between $\UniWqoA_1$ and $\UniWqoA_2$ is a mapping
$\UniReflectionA :  \UniWqoSet_1 \to \UniWqoSet_2$
such that,
for all $\UniValA, \UniValB \in \UniWqoSet_1$,
$
\text{if } \UniReflectionA(\UniValA) \UniWqoRel{2} \UniReflectionA(\UniValB), \text{then } \UniValA \UniWqoRel{1} \UniValB
$
and, for all $a \in \UniWqoSet_1$,
$
\UniNorm{\UniReflectionA(a)}{2} 
\leq \UniNorm{a}{1}.
$
We
write $\UniWqoA_1 \UniReflArrow{\UniReflectionA} \UniWqoA_2$
(sometimes omitting the superscript $\UniReflectionA$).
\end{defn}

\begin{lemmaapx}[{\cite[p.~446]{schmitz2011}}]\label{lem:ref-transf-res}
    Whenever
    $\UniWqoA_1 \UniReflArrow{} \UniWqoA_2$,
            if $\UniWqoA_2$ is a nwqo, then so is $\UniWqoA_1$; and
            if $\UniControlFunctionA$ is a control function, then $\UniLengFunc{\UniWqoA_1}{\UniControlFunctionA}(t) \leq \UniLengFunc{\UniWqoA_2}{\UniControlFunctionA}(t)$
    for all $t \in \UniNaturalSet$.
\end{lemmaapx}
   
\subsection{A well-quasi-order over sequents based on weakening}

Observe that the antecedent of the premise of the left-weakening rule $\UniIRule$ is obtainable from the antecedent of the conclusion by deleting letters, and the succedents are identical. This motivates an nwqo over sequents, for which
we obtain a length theorem 
via a reflection into a disjoint sum of Higman's orderings. 
In what follows, let $\Sigma$ be an arbitrary signature.

\begin{defn}\label{def:wqo-sequents}
    For finite
    $\UniSubfmlaHyperseqSet \subseteq \UniLangSet{\Sigma}{\UniPropSet}$,
    let
    $\UniWqoSeqName{\Sigma}{\UniSubfmlaHyperseqSet}
    \UniSymbDef
    \UniStruct{
    \UniSeqSet{\Sigma}{\UniSubfmlaHyperseqSet},
    \UniSeqWknWqo{\Sigma}{\UniSubfmlaHyperseqSet},
    \UniNormSNCSeq{\cdot}
    }$, where
    \begin{itemize}
        \item $\UniSeqSet{\Sigma}{\UniSubfmlaHyperseqSet}$
        is the set of all $\UniSubfmlaHyperseqSet$-sequents
        over $\UniLangSet{\Sigma}{\UniPropSet}$;
        \item $\UniSeqWknWqo{\Sigma}{\UniSubfmlaHyperseqSet} 
        \;\subseteq \UniSeqSet{\Sigma}{\UniSubfmlaHyperseqSet} \times \UniSeqSet{\Sigma}{\UniSubfmlaHyperseqSet}$
        is such that
        $\UniSequentA_1 \UniSeqWknWqo{\Sigma}{\UniSubfmlaHyperseqSet} \UniSequentA_2$
        iff $\UniSequentA_2$ is obtained by 
        successive applications of $\UniIRule$
        starting from $\UniSequentA_1$;
        \item $\UniNormSNCSeq{\cdot} : \UniSeqSet{\Sigma}{\UniSubfmlaHyperseqSet} \to 
        \UniNaturalSet$
        is such that
        $\UniNormSNCSeq{\UniSequent{\UniMSetFmA}{\UniMSetSucA}} \UniSymbDef
        \UniListLength{\UniMSetFmA}$
        (i.e., the length of the antecedent).
    \end{itemize}
\end{defn}

\begin{theoremapxrep}\label{the:qo-seq-is-nwqo}
    For all finite $\UniSubfmlaHyperseqSet \subseteq \UniLangSet{\Sigma}{\UniPropSet}$,
            $\UniWqoSeqName{\Sigma}{\UniSubfmlaHyperseqSet}$
    is a nwqo; and
            if $\UniControlFunctionA$ is primitive recursive, then
        $\UniLengFunc{\UniWqoSeqName{\Sigma}{\UniSubfmlaHyperseqSet}}{\UniControlFunctionA}$
        is upper bounded by a function in
            $\UniGrzHLevel{\omega^{\UniSetCard{\UniSubfmlaHyperseqSet}-1}}$.
\end{theoremapxrep}
\begin{proof}
    Fix an enumeration
    $\UniFmA_1,\ldots,\UniFmA_m$
    of the formulas in $\UniSubfmlaHyperseqSet$.
    Then the mapping
    $f : \UniSeqSet{\Sigma}{\UniSubfmlaHyperseqSet}
    \to (\UniSetCard{\UniSubfmlaHyperseqSet} + 1) \cdot \UniSubfmlaHyperseqSet^\ast$
    such that
    $f(\UniSequent{\UniMSetFmA}{})
    \UniSymbDef (0, \UniMSetFmA)$
    and
    $f(\UniSequent{\UniMSetFmA}{\UniFmA_j})
    \UniSymbDef (j, \UniMSetFmA)$
    is a reflection; hence the result follows
    from
    Thm.~\ref{thm:hig-leng-the}
    and
    Lem.~$\ref{lem:ref-transf-res}$.
\end{proof}


\subsection{Decision procedure for amenable sequent calculi}

Fix an amenable sequent calculus $\UniSeqCalcA$ (see Def.~\ref{def:amenable-seq-calc})
over a signature $\Sigma$.

\begin{defn}
    For $\mathcal{X}$ a
    formula, a sequent, or a sequent rule,
    let $\UniSizeHyper{\mathcal{X}}$
    be the length of the written representation of
    $\mathcal{X}$.
    If $X$ is a finite set of formulas,
    of sequents, or of sequent rules, let 
    $\UniSizeHyperSum{X} \UniSymbDef \sum_{\mathcal{X} \in X} \UniSizeHyper{\mathcal{X}}$
    and
    $\UniSetSizeMax{X} \UniSymbDef \max_{\mathcal{X} \in X} \UniSizeHyper{\mathcal{X}}$.
    In particular,
    $\UniSetSizeMax{\UniSeqCalcA} \UniSymbDef \max_{\UniRuleA \in \UniSeqCalcA} \UniSizeHyper{\UniRuleA}$ i.e., the largest among the representation lengths of the rules in the calculus.
\end{defn}
The construction in this subsection extends the commutative setting that appears in Balasubramanian et al.~\cite{BalLanRam21LICS}.
For a finite set $\UniDerivSet$ of sequents,
define
    $\UniNormSNCSeqSet{\UniDerivSet} 
    \UniSymbDef
    \max_{\UniSequentA \in \UniDerivSet} \UniNormSNCSeq{\UniSequentA}$
    ($\UniNormSNCSeq{\UniSequentA}$
    is the length of the antecedent of $\UniSequentA$, see Def.~\ref{def:wqo-sequents}).

\begin{defn}\label{def-derive-sets-hfli}
Let $\UniSubfmlaHyperseqSet$
be a finite set of formulas
closed under subformulas
and $\mathcal{T}$ be a regular theory containing only
$\UniSubfmlaHyperseqSet$-sequents.
Define $\UniDerivSet_{0}$ as the set of minimal elements with respect to $\UniSeqWknWqo{\Sigma}{\UniSubfmlaHyperseqSet}$ (i.e., not obtainable from another element by repeated $\UniIRule$) from the following finite set: the union of $\mathcal{T}$
with
the set of
all instances of initial sequents 
in~$\UniSeqCalcA$ that satisfy the following:
\begin{enumerate}[a)]
\item formula-variables are instantiated to elements of~$\UniSubfmlaHyperseqSet$;
\item succedent-variables are instantiated to an element in~$\UniSubfmlaHyperseqSet$ or as empty; 
\item sequence-variables are instantiated to 
the empty sequence.
\end{enumerate}

\noindent Define
$\UniDerivSet_{i+1}:=\UniDerivSet_i \cup \partial\UniDerivSet_i$ ($i\geq 0$),
where $\partial\UniDerivSet_i$
is the set of $\UniSubfmlaHyperseqSet$-sequents
$\UniSequentA$
satisfying:
\begin{enumerate}
    \item\label{conI} $\UniSequentA_1 \cdots \UniSequentA_k/\UniSequentA$ is a rule instance of $\UniSeqCalcA$ such that,
     for all $1 \leq j \leq k$,
    there is $\minus{\UniSequentA_j}\in \UniDerivSet_i$
    with $\minus{\UniSequentA_j}\UniSeqWknWqo{\Sigma}{\UniSubfmlaHyperseqSet}~{\UniSequentA_j}$;
    \item\label{conII}
    the antecedent
    of $\UniSequentA$
    has length $\leq
    (\UniSizeHyper{\UniSeqCalcA}\UniNormSNCSeqSet{\UniDerivSet_i})
    \UniMulttNumbers \UniSizeHyper{\UniSeqCalcA}$;
    \item\label{conIII} there is no $\UniSequentA'\in \UniDerivSet_i$ such that $\UniSequentA'\UniSeqWknWqo{\Sigma}{\UniSubfmlaHyperseqSet} \UniSequentA$.
\end{enumerate}
\end{defn}

The above construction yields a chain $D_0\subseteq D_1\subseteq\ldots$ of sets of sequents. The following shows that the chain stabilizes at a finite index. Indeed, if that were not the case, Def.~\ref{def-derive-sets-hfli} (\ref{conIII}) forces the existence of an infinite bad sequence $(s_i)_{i<\omega}$ from $s_i\in D_i\setminus D_{i-1}$ ($i < \omega$), and that is impossible.

\begin{theoremapxrep}
    \label{the:termination-alg-flw}
    $D_{i+1}$ is computable from $D_i$
    and
    there is $N \in \UniNaturalSet$
    such that $\UniDerivSet_N = \UniDerivSet_{N+i}$
    for all $i \geq 0$.
\end{theoremapxrep}
\begin{proof}
    Each  $\UniDerivSet_i$ is
    computable because 
    (a) there are only finitely-many rules in $\UniSeqCalcA$,
    (b) finitely-many formulas in $\UniSubfmlaHyperseqSet$,
    (c) Def. \ref{def-derive-sets-hfli} (\ref{conII}) restricts the length of antecedents
    (thus there are only finitely-many rule instances to consider),
    and (d) the relation $\UniSeqWknWqo{\Sigma}{\UniSubfmlaHyperseqSet}$ is computable.
    
    If there is some $N$ such that $D_N=D_{N+1}$ then, by the construction of these sets, $D_N=D_{N+i}$ for every~$i\geq 0$. Suppose that
    no such $N$ exists. It follows that
    $D_0 \subset D_1 \subset D_2 \subset \ldots$. Choose any $\UniSequentA_i\in D_i\setminus D_{i-1}$ for each $i\in\mathbb{N}$. By Def. \ref{def-derive-sets-hfli} (\ref{conIII}),
    $(\UniSequentA_i)_{i<\omega}$
    is an infinite bad sequence over 
    $\UniWqoSeqName{\Sigma}{\UniSubfmlaHyperseqSet}$, contradicting that the latter is a wqo (Thm.~\ref{the:qo-seq-is-nwqo}).
    \qedhere
\end{proof}

We now establish (Thm.~\ref{the:dec-alg-flw}) that every deducible sequent is obtainable by weakening from a sequent at the stabilization point. First, a technical lemma.

\begin{lemmaapxrep}
    \label{lem:break-list-higman}
    If
    ${\UniSequentA'}  \UniSeqWknWqo{\Sigma}{\UniSubfmlaHyperseqSet} \UniSequentA$
    with $\UniSequentA = \UniSequent{L_1\cdots L_m}{\UniMSetSucA}$, then
    there are 
    sequences
    $L'_1,\ldots,L'_m$
    such that
    $\UniSequentA' = \UniSequent{L'_1\cdots L'_m}{\UniMSetSucA}$
    and
    $L'_i \UniWqoRel{\ast} L_i$
    for all $1 \leq i \leq m$.
\end{lemmaapxrep}
\begin{proof}
    By induction on a deduction
    $\delta$ witnessing ${\UniSequentA'}  \UniSeqWknWqo{\Sigma}{\UniSubfmlaHyperseqSet} \UniSequent{L_1\cdots L_m}{\UniMSetSucA}$.
    If it has a single node,
    the involved sequents are the same,
    so take $L'_i \UniSymbDef L_i$.
    In the inductive step,
    assume
    $s = \UniSequent{L_1 \cdots L^1_i \UniFmA L^2_i \cdots L_m}{\UniMSetSucA}$
    was obtained by weakening from 
    $s'' = \UniSequent{L_1 \cdots L^1_i L^2_i \cdots L_m}{\UniMSetSucA}$,
    where $L_i = L^1_i L^2_i$.
    Then, by the (IH), $s' = \UniSequent{L'_1 \cdots L'_i \cdots L'_m}{\UniMSetSucA}$
    with $L'_j \UniWqoRel{\ast} L_j$
    for all $1 \leq j \leq m$.
    In particular,
    $L'_i \UniWqoRel{\ast} L_i^1L_i^2 \UniWqoRel{\ast} L_i^1\UniFmA L_i^2$, and we are done.
\end{proof}

\begin{theoremapx}
\label{the:dec-alg-flw}
Let $\mathcal{T} \cup \{ \UniSequentA \}$
be a finite set of $\UniSubfmlaHyperseqSet$-sequents
    such that $\mathcal{T}$ is a regular theory (as in Def.~\ref{def-derive-sets-hfli}).
Then
$\mathcal{T}\UniHyperDerivRel{\UniSeqCalcA} \UniSequentA$ iff
there is $M \in \UniNaturalSet$
such that
$\UniSequentA' \UniSeqWknWqo{\Sigma}{\UniSubfmlaHyperseqSet} 
\UniSequentA$ for some $\UniSequentA' \in \UniDerivSet_M$.
\end{theoremapx}
\begin{proof}
    The left-to-right direction is the non-trivial one.
    Since $\UniSeqCalcA$ is amenable, $\mathcal{T}\UniHyperDerivRel{\UniSeqCalcA} \UniSequentA$ is
    witnessed by an analytic deduction $\delta$.
    We argue by induction on the structure of $\delta$.
    
    \textit{Base case}. Then $\delta$ is a single node, so the node is a sequent in $\mathcal{T}$ or an instance
    of an initial sequent. So $\UniSequentA$ is in $\UniDerivSet_0$, or $\UniSequentA$ is obtainable from the node by applications of
    $\UniIRule$.

    \textit{Inductive step}. The last rule
    instance in $\delta$ is an instance $\UniSequentA_1\cdots\UniSequentA_k/\UniSequentA$ of $\UniRuleA$.
    Since each $\UniSequentA_i$ is an
    $\UniSubfmlaHyperseqSet$-sequent
    by virtue of the deduction being analytic, by applying IH for each $1 \leq i \leq k$, there is $N_i \in \UniNaturalSet$ with ${\UniSequentA'_i}  \UniSeqWknWqo{\Sigma}{\UniSubfmlaHyperseqSet} \UniSequentA_i$
    for some ${\UniSequentA'_i} \in \UniDerivSet_{N_i}$.
    Let $M \UniSymbDef \max_i N_i$,
    which gives us $\UniSequentA'_i \in \UniDerivSet_{M}$
    for all $1 \leq i \leq k$.
    Hence Def.~\ref{def-derive-sets-hfli} (\ref{conI}) is satisfied
    for $\UniSequentA$. The situation is the following.
    
    \begin{center}
        \AxiomC{$\quad\qquad\UniSequentA'_1 \in D_M$}
        \noLine
        \UnaryInfC{\rotatebox{-90}{$\UniSeqWknWqo{\Sigma}{\UniSubfmlaHyperseqSet}$}}
        \noLine
        \UnaryInfC{$\UniSequentA_1$}
        \AxiomC{$\;\cdots$}
        \noLine
        \UnaryInfC{\phantom{\rotatebox{-90}{$\UniSeqWknWqo{\Sigma}{\UniSubfmlaHyperseqSet}$}}}
        \noLine
        \UnaryInfC{$\;\cdots$}
        \AxiomC{$\quad\qquad\UniSequentA'_k \in D_M$}
        \noLine
        \UnaryInfC{\rotatebox{-90}{$\UniSeqWknWqo{\Sigma}{\UniSubfmlaHyperseqSet}$}}
        \noLine
        \UnaryInfC{$\UniSequentA_k$}
        \RightLabel{$\UniRuleA$}
        \TrinaryInfC{$\UniSequentA$}
        \DisplayProof
    \end{center}

%
%
%

    We now show by cases that
    there is $\UniSequentA' \in \UniDerivSet_{M+1}$
    such that
    $\UniSequentA' 
    \UniSeqWknWqo{\Sigma}{\UniSubfmlaHyperseqSet}
    \UniSequentA$.

    If Def.~\ref{def-derive-sets-hfli} (\ref{conII}) is satisfied, then $\UniSequentA\in \UniDerivSet_{M+1}$, or (due to Def.~\ref{def-derive-sets-hfli} (\ref{conIII})) there is some $\UniSequentA' \in \UniDerivSet_{M+1}$
    such that
    $\UniSequentA' 
    \UniSeqWknWqo{\Sigma}{\UniSubfmlaHyperseqSet}
    \UniSequentA$. In either case we are done, so assume that Def.~\ref{def-derive-sets-hfli} (\ref{conII}) fails.
    Thus $\UniSequentA$ has antecedent
    of length $>
    (\UniSizeHyper{\UniSeqCalcA}\UniNormSNCSeqSet{\UniDerivSet_M})
    \UniMulttNumbers \UniSizeHyper{\UniSeqCalcA}$.
    Thus there are sequence-variables 
    $\UniMSetFmA_1,\ldots,\UniMSetFmA_m$ ($m>0$)
    in the conclusion of $\UniRuleA$
    instantiated with
    sequences of length 
    $
    > 
    \UniSizeHyper{\UniSeqCalcA}
    \UniNormSNCSeqSet{\UniDerivSet_M}
    $ (if $m=0$ then the antecedent would be instantiations of formula-variables and hence its length would be $\leq \UniSizeHyper{\UniSeqCalcA}\UniNormSNCSeqSet{\UniDerivSet_0}$).
    Assume wlog that $\UniMSetFmA_1,\ldots,\UniMSetFmA_{m'}$
    with $m'\leq m$
    appear in the premises of $\UniRuleA$,
    while the other variables appear
    exclusively in the conclusion.

    Write 
    $\UniSequentA_i = \UniSequent{L_{i1}\cdots L_{iu_i}}{\UniMSetSucA_i}$
    and
    $\UniSequentA_i' = \UniSequent{L_{i1}'\cdots L'_{i{u_i}}}{\UniMSetSucA_i}$
    where each $L_i$ is an instantiation
    of a sequence-variable or
    a singleton sequence of a formula, such that
    $L'_{ij} \UniWqoRel{\ast} L_{ij}$
    for each $1 \leq j \leq u_i$
    (the existence of such sequences
    is guaranteed by Lem.~\ref{lem:break-list-higman}).
    The length of
    $L_{i1}'\cdots L_{iu_i}'$ is $\leq \UniNormSNCSeqSet{\UniDerivSet_M}$
    by the fact that $\UniSequentA_i' \in \UniDerivSet_M$, 
    hence each $L_{ij}'$ has length
    $\leq \UniNormSNCSeqSet{\UniDerivSet_M}$.

    We construct a new instance $I'$
    of $\UniRuleA$, with premises $\UniSequentA''_i$ instead of $\UniSequentA_i$, and a conclusion $\UniSequentA'$ whose antecedent has size satisfying Def.~\ref{def-derive-sets-hfli}(\ref{conII}), and $\UniSequentA' 
    \UniSeqWknWqo{\Sigma}{\UniSubfmlaHyperseqSet}
    \UniSequentA$:
    \begin{center}
        \AxiomC{$\quad\qquad\UniSequentA'_1 \in D_M$}
        \noLine
        \UnaryInfC{\rotatebox{-90}{$\UniSeqWknWqo{\Sigma}{\UniSubfmlaHyperseqSet}$}}
        \noLine
        \UnaryInfC{$\UniSequentA''_1$}
        \AxiomC{$\;\cdots$}
        \noLine
        \UnaryInfC{\phantom{\rotatebox{-90}{$\UniSeqWknWqo{\Sigma}{\UniSubfmlaHyperseqSet}$}}}
        \noLine
        \UnaryInfC{$\;\cdots$}
        \AxiomC{$\quad\qquad\UniSequentA'_k \in D_M$}
        \noLine
        \UnaryInfC{\rotatebox{-90}{$\UniSeqWknWqo{\Sigma}{\UniSubfmlaHyperseqSet}$}}
        \noLine
        \UnaryInfC{$\UniSequentA''_k$}
        \RightLabel{$\UniRuleA$}
        \TrinaryInfC{\qquad\qquad\qquad\quad\;\;$\UniSequentA'$
        (smaller than $s$)}
        \DisplayProof
    \end{center}
    Construct $I'$ by consideration of where each $\UniMSetFmA_j$ ($1 \leq j \leq m'$) occurs.
    For $\UniMSetFmA_j$  occurring in a single premise, instantiate it
    with the sequence $L'_{jl}$
    instead of the $L_{jl}$ that was used before.
    
    For each $\UniMSetFmA_j$ ($1 \leq j \leq m'$) occurring 
    in multiple premises, we need a single instantiation that can be used for each occurrence (reflecting the fact that we are dealing with additive rules here).
    Without loss of generality, assume that
    $\UniSequentA_1, \ldots,\UniSequentA_l$
    are all the premises where
    $\UniMSetFmA_j$ occurs.
        Assume that $\UniMSetFmA_j$
    was instantiated with the sequence $L$.
    Let $L_{1{j_1}},\ldots,L_{l{j_{l}}}$
    be the sequences in these premises that
    instantiate $\UniMSetFmA_j$ (each is equal
    to $L$).
    We want a subsequence of $L$ reachable
    by weakening from each $L'_{1{j_1}},\ldots,L'_{l{j_{l}}}$ (these need not be identical):
    start with $L$ and cross off each of the sequences
    $L'_{1{j_1}},\ldots,L'_{l{j_{l}}}$
    (letter by letter, respecting the order).
    Use the sequence crossed-off to instantiate $\UniMSetFmA_j$. It has size
    $\leq k \cdot \UniNormSNCSeqSet{\UniDerivSet_M} < \UniSizeHyper{\UniSeqCalcA} \cdot \UniNormSNCSeqSet{\UniDerivSet_M}$.
    
    Finally, for sequence-variables that
    appear only in the conclusion
    of the rule, i.e.,
    $\UniMSetFmA_{m'+1},\ldots,\UniMSetFmA_m$, 
    instantiate each of these with
    the empty sequence.
    
    The new instantiation
    $I'$ has
    premises $\UniSequentA''_i$
    where $\UniSequentA'_i
     \UniSeqWknWqo{\Sigma}{\UniSubfmlaHyperseqSet} \UniSequentA''_i$
     for each $1 \leq i \leq k$, and its conclusion is $\UniSequentA' \UniSeqWknWqo{\Sigma}{\UniSubfmlaHyperseqSet} \UniSequentA$ since the instantiation of each $\UniMSetFmA_j$ is a subsequence of the original instantiation. Moreover, the number of variables in the conclusion of $\UniRuleA$ is $<\UniSizeHyper{\UniSeqCalcA}$ and each variable instantiation under $I'$ has size $<\UniSizeHyper{\UniSeqCalcA} \UniNormSNCSeqSet{\UniDerivSet_M}$. Hence it follows that $\UniSequentA'\in \UniDerivSet_{M+1}$ or there is some $\UniSequentA'_0\in \UniDerivSet_{M}$ with $\UniSequentA'_0\UniSeqWknWqo{\Sigma}{\UniSubfmlaHyperseqSet} \UniSequentA'$. In either case, the claim is proved.\qedhere
\end{proof}


\subsection{Complexity analysis}

The fast-growing complexity classes are closed under exponentiation so the distinction
between space and time (also determinism and
non-determinism) is unimportant~\cite{schmitz2016hierar}.
We undertake a space analysis of the algorithm based on Thm.~\ref{the:termination-alg-flw}
and Thm.~\ref{the:dec-alg-flw}.
It suffices to write down each $\UniDerivSet_i$ in turn at the same location, computing each rule instance in turn; when the stabilization index~$\UniDerivSet_N$ is reached, check whether
$\UniSequentA' \UniSeqWknWqo{\Sigma}{\UniSubfmlaHyperseqSet} \UniSequentA$
for some $\UniSequentA' \in \UniDerivSet_N$.
Each $\UniDerivSet_i \subseteq \UniDerivSet_N$, and writing down a rule instance takes space bounded by the size of an element from $\UniDerivSet_N^{\UniSizeHyper{\UniSeqCalcA}} \times \UniDerivSet_N$.
So the space requirements are an elementary function in $\UniSetCard{\UniDerivSet_N}$ and the size of an element from $\UniDerivSet_N$.
We upper bound each of these.



\begin{lemmaapxrep}\label{lem:ub-sizes}
    Let $\mathcal{T}$ be a regular theory, $\UniSequentA$ be a sequent
    and $\UniSubfmlaHyperseqSet \UniSymbDef 
    \UniSubf{\mathcal{T} \cup \{ \UniSequentA \}}$.
    
    \begin{enumerate}[1.]
        \item $\UniSetCard{\UniSubfmlaHyperseqSet} \leq \UniSizeHyper{\UniSequentA} + \UniSizeHyperSum{\mathcal{T}}$;
        \item if 
        $\UniSequentA' \in \UniDerivSet_i$
        then $\UniNormSNCSeq{\UniSequentA'} < 
        (\UniSizeHyperSum{\mathcal{T}}+1) \UniSizeHyper{\UniSeqCalcA}^{2i+1}$,
        and $\UniNormSNCSeqSet{\UniDerivSet_i} < 
        (\UniSizeHyperSum{\mathcal{T}}+1)\UniSizeHyper{\UniSeqCalcA}^{2i+1}$;
        \item if 
        $\UniSequentA' \in \UniDerivSet_i$
        then $\UniSizeHyper{\UniSequentA'} 
        < 4  (\UniSizeHyperSum{\mathcal{T}} + \UniSizeHyper{\UniSequentA})^2 \UniSizeHyper{\UniSeqCalcA}^{2N+1}$;
        \item 
        for all $i \geq 0$,
       $\UniSetCard{\UniDerivSet_i} < E(\UniSizeHyperSum{\mathcal{T}}+\UniSizeHyper{\UniSequentA}, N)$
        for $E$ an elementary function (i.e., in $\UniGrzHLevel{2}$).
    \end{enumerate}
\end{lemmaapxrep}
\begin{proof}
    \begin{enumerate}[1.]
        \item It is easy to see by
        structural induction that
        $\UniSetCard{\UniSubf{\UniFmA}} \leq \UniSizeHyper{\UniFmA}$ for all formulas $\UniFmA$.
        Then, for any sequent $\UniSequentB \UniSymbDef \UniSequent{\UniFmB_1,\ldots,\UniFmB_m}{\UniMSetSucA}$,
        we have
        $\UniSetCard{\UniSubf{\UniSequentB}} \leq \UniSetCard{\UniSubf{\UniMSetSucA}} + \sum_{i=1}^m\UniSetCard{\UniSubf{\UniFmB_i}}
        \leq \UniSizeHyper{\UniMSetSucA} + \sum_{i=1}^m \UniSizeHyper{\UniFmB_i} \leq \UniSizeHyper{\UniSequentB}$. 
        The result then follows because
        $\UniSetCard{\UniSubfmlaHyperseqSet} \leq 
        \UniSetCard{\UniSubf{\UniSequentA}}
        + \sum_{\UniSequentB \in \mathcal{T}}
        \UniSetCard{\UniSubf{\UniSequentB}}
        \leq\UniSizeHyper{\UniSequentA} + \UniSizeHyperSum{\mathcal{T}}
        $.
        \item 
        By induction on $i \geq 0$.
        Note that
        if $\UniSequentA' \in \UniDerivSet_0$, since $D_0$ is a subset of the union of $\mathcal{T}$ and instantiations from $\UniSubfmlaHyperseqSet$ of initial sequents in $\UniSeqCalcA$,
        $\UniNormSNCSeq{\UniSequentA'} < \UniNormSNCSeqSet{\mathcal{T}} + \UniSizeHyper{\UniSeqCalcA} \leq (\UniNormSNCSeqSet{\mathcal{T}}+1) \UniSizeHyper{\UniSeqCalcA}
        \leq (\UniSizeHyperSum{\mathcal{T}}+1)\UniSizeHyper{\UniSeqCalcA}$.
        Inductive step:
        if $\UniSequentA' \in \UniDerivSet_{i+1}$,
        then $\UniNormSNCSeq{\UniSequentA'} < \UniSizeHyper{\UniSeqCalcA}^2\UniNormSNCSeqSet{\UniDerivSet_i}
        <
        \UniSizeHyper{\UniSeqCalcA}^2(\UniSizeHyperSum{\mathcal{T}}+1)\UniSizeHyper{\UniSeqCalcA}^{2i+1} = (\UniSizeHyperSum{\mathcal{T}}+1)\UniSizeHyper{\UniSeqCalcA}^{2(i+1)+1}$.

        \noindent Since $\UniNormSNCSeqSet{\UniDerivSet} 
    =
    \max_{\UniSequentA \in \UniDerivSet} \UniNormSNCSeq{\UniSequentA}$, it follows that $\UniNormSNCSeqSet{\UniDerivSet_i} < 
        (\UniSizeHyperSum{\mathcal{T}}+1)\UniSizeHyper{\UniSeqCalcA}^{2i+1}$.
        \item 
        It is enough to observe that
        $\UniSizeHyper{s'}$
        is upper bounded by
        the sum of the sizes of each formula
        in the antecedent, plus 1 for each comma,
        plus 1 to account for the sequent symbol,
        plus the size of the succedent.
        This gives us
        $\UniSizeHyper{s'} \leq \UniNormSNCSeq{s'}\UniSizeHyper{\UniSubfmlaHyperseqSet}
        + \UniNormSNCSeq{s'} + 1 + \UniSizeHyper{\UniSubfmlaHyperseqSet}
        \leq 4(\UniNormSNCSeq{s'}+1)\UniSizeHyper{\UniSubfmlaHyperseqSet}$. From the previous items, the latter is 
$\leq 4((\UniSizeHyperSum{\mathcal{T}}+1)\UniSizeHyper{\UniSeqCalcA}^{2N+1}+1)(\UniSizeHyper{\UniSequentA} + \UniSizeHyperSum{\mathcal{T}})
\leq
4((\UniSizeHyperSum{\mathcal{T}}+\UniSizeHyper{\UniSequentA})\UniSizeHyper{\UniSeqCalcA}^{2N+1})(\UniSizeHyper{\UniSequentA} + \UniSizeHyperSum{\mathcal{T}})$ where we have used that $\UniSizeHyper{\UniSequentA}\geq 2$ and $\UniSizeHyper{\UniSeqCalcA}\geq 1$.

        \item 
            Let $\eta,\rho \in \UniNaturalSet$.
            Then define
         $\UniNDistCompNC{\eta}{\rho} \UniSymbDef (\sum_{l=0}^{\rho} \eta^l) \cdot (\eta+1) \leq (\eta+1)^{\rho+1}\cdot (\eta+1)$, the number of distinct sequents
        over $\eta$ formulas and having norm
        $\leq\rho$.
        Note that
        $\UniSetCard{\UniDerivSet_0} \leq \UniNDistCompNC
        {\UniSizeHyper{\UniSequentA} + \UniSizeHyperSum{\mathcal{T}}}
        {\UniSizeHyper{\UniSequentA} + \UniSizeHyperSum{\mathcal{T}}}$
        and
        $\UniSetCard{D_{i+1}} 
        \leq 
        \UniNDistCompNC{\UniSetCard{\UniSubfmlaHyperseqSet}}
        {\UniSizeHyper{\UniSeqCalcA}^2\UniNormSNCSeqSet{D_i}}
        < 
        \UniNDistCompNC{\UniSizeHyper{\UniSequentA} + \UniSizeHyperSum{\mathcal{T}}}
        {\UniSizeHyper{\UniSeqCalcA}^2(\UniSizeHyperSum{\mathcal{T}+1)}\UniSizeHyper{\UniSeqCalcA}^{2N+1}}
        \leq 
        \UniNDistCompNC{\UniSizeHyper{\UniSequentA} + \UniSizeHyperSum{\mathcal{T}}}
        {\UniSizeHyper{\UniSeqCalcA}^2(\UniSizeHyper{\UniSequentA} + \UniSizeHyperSum{\mathcal{T}})\UniSizeHyper{\UniSeqCalcA}^{2N+1}}
        $. Thus,
        for all $i \geq 0$,
        $\UniSetCard{\UniDerivSet_i} < E(\UniSizeHyper{\UniSequentA}+\UniSizeHyperSum{\mathcal{T}}, N)$
        for $E$ an elementary function (i.e. in $\UniFGHLevel{2}$).\qedhere
    \end{enumerate}
\end{proof}


\begin{theoremapx}\label{the:complexity-result}
    Deciding whether
    $\mathcal{T}\UniHyperDerivRel{\UniSeqCalcA}\UniSequentA$
    for a regular theory $\mathcal{T}$
    is in 
    $\UniHAck$.
\end{theoremapx}
\begin{proof}
Lem.~\ref{lem:ub-sizes} (4) shows that $\UniSetCard{\UniDerivSet_N} < 
E_1(\UniSizeHyperSum{\mathcal{T}}+\UniSizeHyper{\UniSequentA}, N)$, for an elementary function $E_1$. Also, Lem.~\ref{lem:ub-sizes} (3) shows that that $\UniSequentA'\in\UniDerivSet_N$ implies that its size $\UniSizeHyper{\UniSequentA'}$ is bounded by $E_2(\UniSizeHyperSum{\mathcal{T}}+\UniSizeHyper{\UniSequentA}, N)$,
for an elementary function $E_2$.
It remains to upper bound $N$, and it is this that forces the fast-growing complexity.
    Let $\UniSubfmlaHyperseqSet \UniSymbDef \UniSubf{\mathcal{T} \cup \{ \UniSequentA \}}$.
    In the proof of Thm.~\ref{the:termination-alg-flw},
    we extracted a bad sequence
    $(\UniSequentA_i)_{i \leq N}$
    over $\UniWqoSeqName{\Sigma}{\UniSubfmlaHyperseqSet}$
    from the chain of sets $D_0 \subset \ldots \subset D_N$ by choosing any $\UniSequentA_i\in D_i\setminus D_{i-1}$. 
    We claim that this bad sequence is $(\UniControlFunctionA,n)$-controlled
    for the control function $\UniControlFunctionA(x) \UniSymbDef \UniSizeHyper{\UniSeqCalcA}^2x$
    and $n \UniSymbDef (\UniSizeHyperSum{\mathcal{T}}+1)\UniSizeHyper{\UniSeqCalcA}$, allowing us to conclude that
    $N \leq \UniLengFunc{\UniWqoSeqName{\Sigma}{\UniSubfmlaHyperseqSet}}{\UniControlFunctionA}(n) \leq \UniLengFunc{\UniWqoSeqName{\Sigma}{\UniSubfmlaHyperseqSet}}{\UniControlFunctionA}((\UniSizeHyperSum{\mathcal{T}}+\UniSizeHyper{\UniSequentA})\UniSizeHyper{\UniSeqCalcA})$.
    By induction on $i$. \emph{Base case}: $\UniSequentA_0 \in \UniDerivSet_0$,
    so Lem.~\ref{lem:ub-sizes}(2) yields $\UniNormSNCSeq{\UniSequentA_0} < (\UniSizeHyperSum{\mathcal{T}}+1)\UniSizeHyper{\UniSeqCalcA}$.
    \emph{Inductive step}:
    $\UniNormSNCSeq{\UniSequentA_{i+1}} < (\UniSizeHyperSum{\mathcal{T}}+1)\UniSizeHyper{\UniSeqCalcA}^{2(i+1)+1} =
    \UniSizeHyper{\UniSeqCalcA}^2(\UniSizeHyperSum{\mathcal{T}}+1)\UniSizeHyper{\UniSeqCalcA}^{2i+1} = \UniSizeHyper{\UniSeqCalcA}^2\UniControlFunctionA^i(\UniSizeHyperSum{\mathcal{T}}+1)\UniSizeHyper{\UniSeqCalcA}) = g^{i+1}(n)$, so we are done.
    So the space to run the algorithm
    is a composition of elementary functions with
    a function in $\UniGrzHLevel{\omega^{\UniSetCard{\UniSubfmlaHyperseqSet}-1}}$ 
    by Thm.~\ref{the:qo-seq-is-nwqo},
    thus it is bounded by a function in
    $\UniGrzHLevel{\omega^{\UniSetCard{\UniSubfmlaHyperseqSet}-1}}$.
    So, for this
    particular $\UniSetCard{\UniSubfmlaHyperseqSet}$,
    the problem is in $\UniFGHProbOneAppLevel{\omega^{\UniSetCard{\UniSubfmlaHyperseqSet}}}$.
    As $\UniSubfmlaHyperseqSet$ varies with the input, we wish to eliminate its dependence. Upper bounding over all $\UniSubfmlaHyperseqSet$, we have that
    the problem is in
    $\UniFGHProbOneAppLevel{\omega^{\omega}} = \UniHAck$.
\end{proof}


\begin{corollaryapx}
    Any subcalculus of $\UniFLExtSCalc{\UniWProp}{}$ 
    containing $\UniIRule$ has deducibility in
    $\UniHAck$.
    This holds in particular for
    $\UniFLExtSCalc{\UniWProp}{\Sigma}$
    for any $\Sigma \subseteq \UniSigFL{}$.
\end{corollaryapx}



\section{Final considerations}
\label{sec:final-considerations}
Combining Sec.~\ref{sec:lower-bounds} and
 Sec.~\ref{sec:upper-bounds}, we finally obtain the promised result:
%
%
\vspace{\topsep}

\noindent\textbf{Theorem 1} (Main theorem)\textbf{.}\emph{
For $\{ \fus,\ld,0,1 \} \subseteq \Sigma \subseteq \UniSigFL{}$,
            deducibility in 
        $\UniFLExtSCalc{\UniWProp}{\Sigma}$
        is $\UniHAck$-complete.
            In particular, 
            $\UniFLExtLogic{\UniWProp}{}$
            and its
            multiplicative fragment
            $\UniLExtLogic{\UniWProp}{}$
            are $\UniHAck$-complete.
}


\paragraph{Complexity of the word problem in integral FL-algebras}
Let $\mathsf{V}$ be an equational class of algebras (a \emph{variety})  over a signature $\Sigma$.
The \emph{word problem of $\mathsf{V}$}~\cite[Sec. 4.4.2]{GalJipKowOno07} asks
whether, given fixed finite sets of
variables $X$ (generators) and equations $E$ over
$\UniLangSet{\Sigma}{X}$,
the quasiequation $\& E {\implies} e$ is valid in $\mathsf{V}$, where $e$ is an equation over $\UniLangSet{\Sigma}{X}$.
Note that deciding this problem allows for one algorithm per pair $(X, E)$, as opposed to deciding the \emph{quasiequational theory of $\mathsf{V}$},
which asks for a single algorithm that applies to every $(X, E)$.
In view of the algebraizability
of $\vdash_{\UniFLExtLogic{\UniWProp}{}}$~\cite[Sec.~2.6]{GalJipKowOno07}
w.r.t. the variety of integral zero-bounded FL-algebras $\mathsf{FL_w}$, our results
imply that the word problem and the quasiequational theory
of $\mathsf{FL_w}$ are $\UniHAck$-complete.
In fact, because 
$\UniPropSetLcs{\UniLcsA}$ (cf. Def.~\ref{def:props-lcs}) is a finite set of generators and $\UniSeqToFormEnc{\UniTheoryLCS{\UniLcsA}}$ (cf. Def.~\ref{def:theory-lcs})
is a finite set of formulas over $\UniLangSet{\UniSigFL{}}{\UniPropSetLcs{\UniLcsA}}$,
we have that the word problem
in this variety is $\UniHAck$-hard.
The quasiequational theory
is in $\UniHAck$ in view of the
proof search procedure in the previous section.
Since the word problem reduces to the quasiequational theory,
we obtain that both problems are $\UniHAck$-complete.
This also applies to integral FL-algebras and integral residuated lattices since
the constant $0$ does not play any essential role in the arguments.

\paragraph{Non-existence of deduction theorem}
Our lower bounds imply that
fragments of
$\UniFLExtSCalc{\UniWProp}{}$
covered by Thm.~\ref{thm:lbs}
have no deduction theorem (DT); else
 deducibility would reduce to
provability, yet provability in these logics
is \UniCompClass{PSpace}.

\paragraph{Upper bound for axiomatic extensions}
The obtained upper-bounds apply to deducibility for $\mathcal{N}_2$-analytic structural rule extensions of $\UniFLExtSCalc{\UniWProp}{}$, and to the corresponding axiomatic extensions (refer to the substructural hierarchy~\cite{CiaGalTer17}).

\paragraph{Undecidability of deducibility in $\UniFLExtSCalc{}{}$}
The encoding
in Sec.~\ref{sec:encoding}
offers a new proof of the undecidability
of the deducibility problem
in $\UniFLExtSCalc{}{}$ (remove lossiness; reachability in channel systems is undecidable).
This might be useful to prove
undecidability of deducibility in some axiomatic extensions of
$\UniFLExtSCalc{}{}$.

\vfill


\bibliographystyle{elsarticle-num} 
\bibliography{cas-refs}





\end{document}